\documentclass[psfig]{mn2e}
\usepackage{graphicx}

\begin{document}

\title[CALIFA Spectroscopy:  Interacting Galaxy NGC 5394]
 {CALIFA Spectroscopy of the Interacting Galaxy NGC 5394 (Arp 84): Starbursts, Enhanced [NII]6584 and Signs of Outflows and Shocks}
\author[N.D. Roche, A. Humphrey, J.M. Gomes, P. Papaderos, P. Lagos, S.F. S\'anchez]
 {Nathan Roche$^1$
 \thanks{nathan.roche@astro.up.pt},
 Andrew Humphrey$^{1}$, Jean Michel Gomes$^{1}$, Polychronis Papaderos$^{1}$,\\\\
 {\LARGE \rm Patricio Lagos$^{1}$, Sebasti\'an F. S\'anchez$^{2}$}
\\\\
$^1$ Instituto de Astrof\'isica e Ci\^encias do Espa\c co, 
Universidade do Porto, CAUP, Rua das Estrelas, 4150-762 Porto, Portugal.\\
$^2$Instituto de Astronom\'\i a, Universidad Nacional Auton\'oma de M\'exico, A.P. 
70-264, 04510, M\'exico, D.F.\\}

\bibliographystyle{unsrt} \bibliographystyle{unsrt}

\date{23 June 2015}

\pagerange{\pageref{firstpage}--\pageref{lastpage}} \pubyear{2015}

\maketitle
 
\label{firstpage}

\begin{abstract} 
 We investigate  the spiral galaxy NGC 5394, which is strongly interacting with the larger spiral NGC 5395 (the pair is Arp 84), using optical integral-field spectroscopy from the CALIFA (Calar Alto Legacy Integral Field Area) survey. Spatially-resolved  equivalent widths, emission-line ratios and kinematics reveal many features related to the interaction, which has reshaped the galaxy. $\rm H\alpha$ maps (with other diagnostic emission lines) show a concentrated central ($r<1$ kpc) starburst and three less luminous star-forming regions (one knot far out in the northern arm), and we estimate the dust-corrected total star-formation rate  as 3.39 $\rm M_{\odot}yr^{-1}$.   However, much of the galaxy, especially the outer tidal arms, has a
  post-starburst spectrum, evidence of a more extensive episode of star-formation a few $\times 10^8$ yr ago, triggered by the previous perigalacticon. The $\rm [NII]6584/H\alpha$ ratio is high in the nucleus, reaching 0.63 at the centre, which we interpret as related to high electron density ($n_e\simeq 750$ $\rm cm^{-3}$ from the $\rm [SII]{6717\over 6731}$ ratio). We find a central region of strong and blueshifted NaI(5890,5896) absorption, indicative of  a starburst-driven outflow from the nucleus at  an estimated velocity $\sim 223$ km $\rm s^{-1}$. The CALIFA data  also show an annular region at radii 2.25--4 kpc from the  nucleus, with elevated ratios of [NII], [OI]6300 etc. to the Balmer lines -- this is evidence of shock excitation, which might be the result of  interaction-triggered gas inflow.
    \end{abstract}
   
\begin{keywords}
 galaxies: individual: NGC 5394, galaxies:interactions, galaxies: evolution, galaxies:starburst
 \end{keywords}

\section{Introduction}
Mergers of two spiral galaxies are one of the most important processes in galaxy evolution, and typically occur in a sequence of stages over several hundred Myr, with successive close passages (perigalactica) of the pair, multiple bursts of star-formation, and complex morphological transformations preceding the final coalescence into a single galaxy (which could become an elliptical or lenticular). Interacting galaxies at any stage can be well studied by integral field spectroscopy (IFS), which may reveal both the current activity and recent history within each resolution element.
 
NGC 5394 and NGC 5395 are an interacting close pair of spirals, which appear to be destined to  merge. This well-known system was included in the Arp (1966) catalog of peculiar galaxies under the single name of Arp 84, and studied further  by Arp (1969). On account of its shape it has also been called the Heron Galaxy, with the larger NGC 5395 being the `body and wings' and the two-armed, northern NGC 5394 the `neck, head and beak'; this is well-depicted on the Gran Telescopio Canarias gallery image  (Fig 1). 

 NGC 5394, a barred spiral seen almost face-on and classed as SB(s)b-pec, appears relatively symmetric, but its structure is unusual and greatly modified by the interaction. Within its compact disk it has a small, bright nucleus and 3 arms (double on the west) and beyond the disk (radii $r>13$ arcsec out to $\sim 40$ arcsec) a distinct and much larger system of two long arms extending north and south, the southern arm reaching NGC 5395.

The Arp 84 system is described in some detail by Kaufman et al. (1999), hereafter K99, who observed in $\rm H\alpha$ with a Fabry-Perot. NGC 5394 and 5395 have a mass and luminosity ratio about 1:4, projected separation 28 kpc, and rotate anticlockwise and  clockwise respectively, so that their arms are trailing.  The interaction (an orbit approximately in the sky plane) is prograde with respect to the rotation of NGC 5394, but retrograde and inclined for NGC 5395.  NGC 5394 has strong $\rm H\alpha$ emission (indicating star-formation) from the nucleus ($2.2\times 10^{-13}$ erg $\rm cm^{-2} s^{-1}$ from Keel et al. 1985) and more from a region in the SW of the disk (the inner western arm) and two bright spots on the inner and outer northern arm. K99's radio observations give the neutral hydrogen (HI) mass in NGC 5394 as $7.3\times10^8\rm M_{\odot}$, and the CO data (tracing molecular hydrogen) show a dense concentration ($>10^9M_{\odot}$) of $\rm H_2$ in the nucleus. K99 describe NGC 5394 as `post-ocular' in that it resembles the `eye-shaped' interacting galaxies IC 2163 and NGC 2535 but is at a slightly later stage where `the eye-shaped oval has evolved into inner spiral arms and most of the gas has fallen into the central region, where it fuels a starburst'.  K99 reported some evidence for  a low-velocity outflow from asymmetry of the $\rm H\alpha$ line (a possible blueshifted component) and small ($\sim 10$ km $\rm s^{-1}$) distortions in the $\rm H\alpha$ velocity map (NE of the nucleus); unconfirmed but  will be investigated here. 

Kaufman et al. (2002), hereafter K02,  performed further CO observations and comparison with several other passbands. They estimate NGC 5394 is inclined $15^{\circ}\pm2$ to the sky plane, discuss evidence (e.g. $\rm 60\mu m$ flux) the $\rm H\alpha$ flux  is reduced by as much as an order of magnitude by internal dust extinction, and on this basis estimate the total (dust-corrected) star-formation rate (SFR) as $6\pm 2\rm~M_{\odot}yr^{-1}$ (with no evidence of an active galactic nuclei - AGN - contribution). The star-formation closely traces the HI and $\rm H_2$ distributions --  all three are concentrated in the central kpc, but also the disk is `lopsided, with more HI, CO, and $\rm H\alpha$ emission coming from the western or southwestern side', and there was speculation that this resulted from a west-vs-east time lag in the formation of disk arms and subsequent infall of gas to the nucleus. They estimate a molecular/atomic ratio $\rm M(H_2)/M(HI)=2.5$--2.7 (even with a lower CO--$\rm H_2$ conversion factor applied for starbursts), very high compared to 0.86 for NGC 5395, which contains much more HI ($>10^{10}\rm M_{\odot}$), and the average ratio of 0.28 for spiral galaxies (Casoli et al. 1998). 
 
Puerari, Valdez-Guti\'errez \& Hern\'andez-L\'opez (2005) studied the morphology of both galaxies in the near-infrared (JHK), and fit (in the $K$-band) a bulge profile with $r_{eff}=3.12$ arcsec plus a disk of scale length $h=6.35$ arcsec for NGC 5394, which is `more compact than normal galaxies'.  From the disturbed, Cartwheel-like structure of NGC 5395 they conclude the system is undergoing a strong interaction, a `collision rather than a grazing encounter'. Kaneko et al. (2013) again study the system in CO and depict the contrast between the compact, nucleated, $\rm H_2$-rich NGC 5394 and the ring-like and HI-rich 5395. 
   \begin{figure}
 \includegraphics[width=0.9\hsize,angle=0]{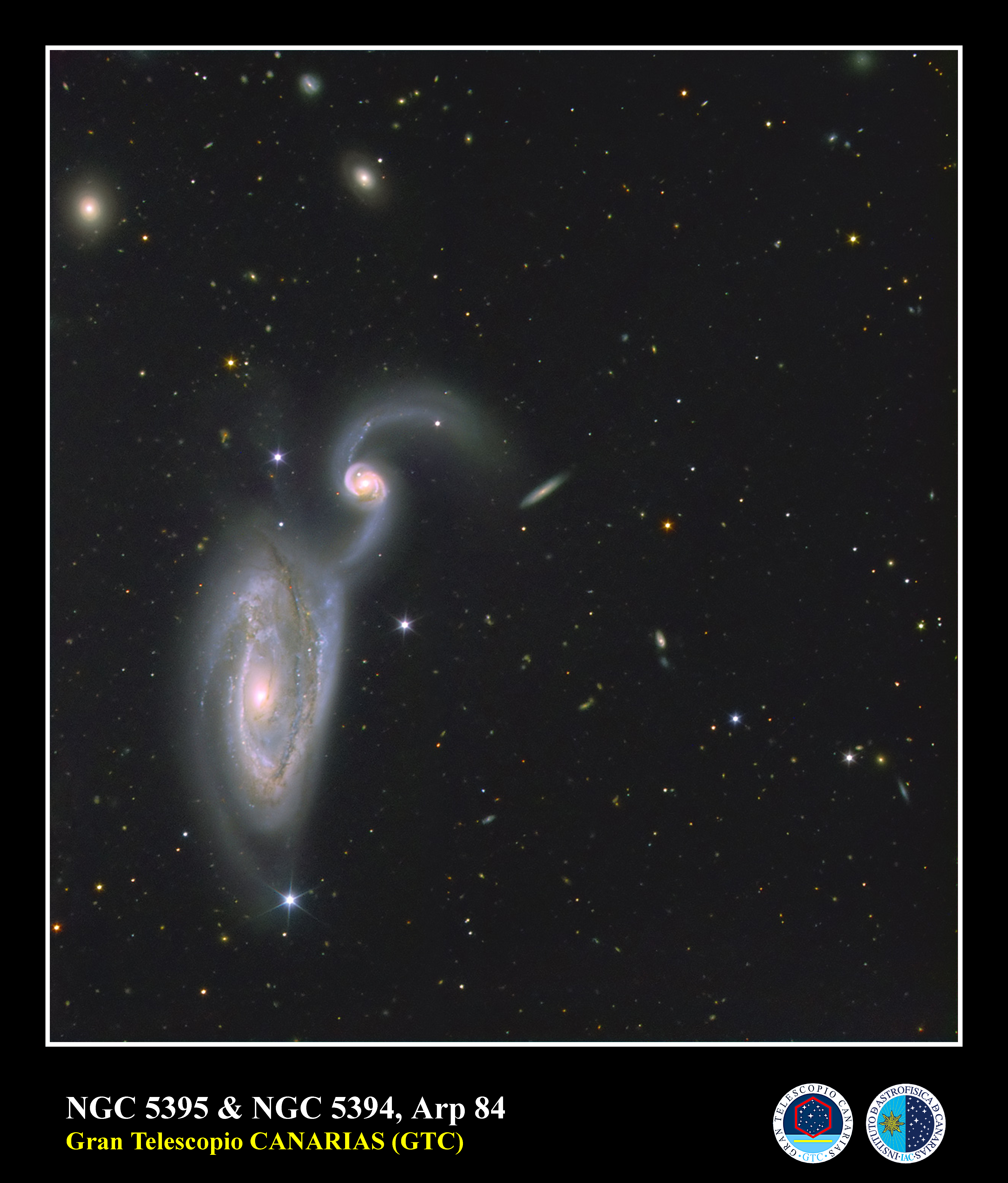}
\caption{Gran Telescopio Canarias public gallery image of Arp 84, NGC 5394 (north) 5395 (south). From 
www.gtc.iac.es/multimedia/media/GTC\_ARP84.jpg} 
 \end{figure}

Smith et al. (2007) observed the system from  3.6 to $24\mu\rm m$ in a {\it Spitzer} survey of Arp-catalog galaxies: most emission from NGC 5394 is from the  compact nucleus  and the {\it Spitzer} images also show the prominent bright spot ($\rm H\alpha$ source) out in the northern arm.  Higdon et al. (2014) describe this hotspot (`N1') as an `intergalactic star-forming object' with a $8\mu\rm m$ flux of $2.626\pm 0.015$ mJy, $\sim 1\%$ that of the whole galaxy ($279.8\pm3.8$ mJy), and in the mid-IR it is also detected in [NeII], [NeIII], [SIII] and polycyclic aromatic hydrocarbons (PAHs). 

 Note that the SFRs estimated from observable quantities will generally depend on the assumed initial mass function (IMF), e.g. the SFR from K02 assumed a Salpeter IMF, whereas equivalent SFR estimates based on the Chabrier or Kroupa IMFs, with fewer low-mass stars, will be 40--$45\%$ lower.
Lehmer et al. (2010) estimated a (Kroupa IMF) SFR of 9.9 $\rm M_{\odot}yr^{-1}$ for the whole Arp 84 system, simply from the infra-red luminosity of $\rm 10^{11}~M_{\odot}yr^{-1}$, and found its X-ray luminosity from {\it Chandra} data consistent with the typical $\rm L_{x}-SFR$ relation of luminous infra-red galaxies (LIRGs), meaning it can be accounted for by X-ray binaries without requiring AGN. Howell et al. (2010) estimated a higher SFR of 20.99 $\rm M_{\odot}yr^{-1}$ by including the unobscured component traced by far-UV luminosity ({\it GALEX}) as well as the infra-red luminosity and modelling with the  Salpeter IMF.

Lanz et al. (2013) combined fluxes in many bands from FUV ({\it GALEX}) to FIR ({\it Herschel}) for a sample of interacting galaxies and fitted the SEDs with a program {\sevensize MAGPHYS}, which assumed a Chabrier IMF and gives a SFR averaged over the past 100 Myr. For  NGC 5394 and 5395, they estimate SFRs of 1.54 and 3.69 $\rm M_{\odot}yr^{-1}$, and found some difference in their SEDs,  with NGC 5394 having relatively more 12-$\rm 60\mu m$ flux and higher  dust temperatures (23.2K in the cold component) in  comparison to 5395 (with 19.7 K) and the non-interacting control sample (about 19K). This is consistent with NGC 5394 experiencing a strong (`stage 4'), and late-stage interaction, with star-formation more centrally concentrated than in the more massive companion. Stellar masses are estimated as 4.79 and $\rm 15.14\times 10^{10}M_{\odot}$.

In this paper we investigate the spatially-resolved properties and history of NGC 5394, especially the effects of the interaction, using new integral field spectroscopy from the Calar Alto Legacy Integral Field Area  (CALIFA) survey 
(S\'anchez et al. 2012; Husemann et al. 2013).  CALIFA has already been the basis  for a number of papers on global properties and on in-depth studies of particular galaxies, with which we can compare -- e.g. the famous `playing Mice'  interacting pair NGC 4676A/B (Wild et al. 2014), and studies of metallicity gradients in typical spirals and mergers (S\'anchez et al. 2014; Barrera-Ballesteros et al. 2015). We will also compare with models of emission-line ratios in star-forming galaxies (e.g. Kewley et al. 2013) and in shocks (Allen et al. 2008); and simulations of galaxy mergers (e.g. Barnes 2004, Struck et al. 2005, Torrey et al. 2012).

From the CALIFA catalog, the J2000 co-ordinates of NGC 5394 are R.A. 13:58:33.19 and Dec +37:27:13.11 and the heliocentric optical recession velocity is 3457 km $\rm ^{-1}$ ($z=0.01153$); the velocity relative to the local standard of rest (Virgo cluster) is given as a slightly greater 3688 km $\rm s^{-1}$ (corresponding to $z=0.01230$). For $\rm H_0=70$ km $\rm s^{-1}Mpc^{-1}$, $\Omega_M=0.27$, $\Omega_{\Lambda}=0.73$ (assumed throughout) the Virgo velocity corresponds to a distance modulus 33.64 mag (the value listed in the CALIFA catalog) and an angular diameter distance 52 Mpc, so that 1 arcsec subtends 252 parsec. In Section 2 of this paper we describe the data, in Section 3 the distribution of star-formation activity, in Section  4 and 5 diagnostic line ratios
 and metallicity, in Section 6 the line-of-sight kinematics and  and Section 7 discuss the history of the interaction (summarizing in Section 8).
  \section{CALIFA Observations} 
 Our data set for NGC 5394 comes from the first CALIFA public data release (DR1; Husemann et al. 2013), which consists of optical datacubes (image $\times$ wavelength) for the first 100 (of what will be eventually 600+)  galaxies in the survey, now updated with Data Release 2 (Garc\'ia-Benito et al. 2015). CALIFA observations are obtained with the integral-field spectrograph Potsdam MultiAperture Spectrophotometer (PMAS)/PPAK mounted on the 3.5m telescope at the Calar Alto observatory.  The PPAK unit contains 331 densely packed optical fibres arranged in a hexagonal area of $74\times 65$ arcsec, to sample an astronomical object at 2.7 arcsec per fibre, with a filling factor of 65\% due to interfibre gaps.

 The CALIFA galaxy sample is selected from the Sloan Digital Sky Survey (SDSS) DR7 catalog, simply on the basis of redshift  ($0.005 < z < 0.03$) and apparent size (angular isophotal diameter $45<D_{25}<80$ arcsec), so that each galaxy fitted within, while filling most of, a single field-of-view. 
This means the survey covers a very wide range of morphological type, absolute magnitude, stellar mass, star-formation rate, colour, etc. (Walcher et al. 2014). 
 NGC 5395 was not targeted in CALIFA as its apparent size is more than twice NGC 5394 (radii $R_{K20}=34$ and 71 arcsec; Kaneko et al. 2013) and exceeds a single field-of-view, so this analysis is limited to the more compact galaxy.

Each galaxy was observed with two spectrograph configurations,  (i) the V500 grating covering the wavelength range 3745--$7500\rm\AA$ with a spectral resolution of $6.0\rm\AA$ (full-width at half-maximum - FWHM), and (ii) the V1200 grating covering 3650--4840$\rm \AA$ with a spectral resolution of $2.3\rm \AA$ (FWHM). Exposure times are 900 sec on each galaxy with V500 and $3\times 600$s with V1200, and the respective $3\sigma$ limits for detection of a narrow emission line are 1.0 and $0.6\times 10^{-17}$ ergs $\rm cm^{-2}s^{-1}$.
  Initial processing of these data generated two (V500 and V1200) data cubes for each galaxy, which  are supplied on the CALIFA website (califa.caha.es) calibrated in units of $\rm \AA$ and $10^{-16}$ erg $\rm cm^{-2}s^{-1}\AA^{-1}$ and rebinned into a grid of 1 arcsec spatial pixels. As the spectrograph fibre diameter is 2.7 arcsec, the effective spatial resolution is lower, with FWHM from 3 to 4 arcsec.
The cubes are $78\times 73$ (spatial) $\times 1877$ (spectral) with $\rm d\lambda=2.0\AA$ for V500, and    $78\times 73\times 1701$ with $\rm d\lambda=0.7\AA$ for V1200.

In this paper we concentrate mostly on the V500 spectra which contain the lines of interest. We have analyzed the datacubes using a locally developed software pipeline called {\sevensize PORTO3D} (Papaderos et al. 2013, Gomes et al. 2015), which effectively separates out the stellar and nebular (emission-line) components of each spatial pixel's spectrum. The stellar component is fitted with a set of model stellar spectra, with a wide range of ages, using the {\sevensize STARLIGHT} package (Cid Fernandes et al. 2013).  The total stellar mass is estimated from the {\sevensize PORTO3D} fit as $4.6\times 10^{10}M_{\odot}$ (for Salpeter IMF).  For a list of spectral lines, the program  generates 2D maps of emission-line flux, equivalent width, velocity offset (red or blueshift), and the equivalent widths for any stellar absorption lines (for Balmer lines $\rm H\beta\gamma\delta$ there may be narrow emission within broader absorption; as was seen for this galaxy by Arp 1969). Further maps show stellar properties estimated from the model fits, such as mean stellar age and metallicity (mass and luminosity weighted). One line map, for [OI]6300, appeared mildly contaminated by the sky line [OI]6364 (de-redshifted to $6291\rm\AA$), so to exclude this a new flux map was computed summing 6297--$6307\rm\AA$ only.

 \section{Mapping the Star-formation}
 Figure 2 shows the CALIFA image of NGC 5394 in the continuum at 6390--$6490\rm\AA$, just blueward of $\rm H\alpha$, where it  appears  near-symmetrical, but the emission lines will reveal a much more complex picture. The $\rm H\alpha$ emission line is the best tracer of the current star-formation rate (SFR). CALIFA maps of  the $\rm H\alpha$  flux and equivalent width (EW, here in the rest-frame) show (Fig 3) the star-formation is concentrated in four distinct regions, the nucleus ($r<4$ arcsec or 1 kpc), a region in the SW of the disk (identified by K99 as the inner of two western spiral arms) and two compact `hotspots' in the inner and outer northern tidal arm (see also Fig 4 of  K99).   Peak $\rm H\alpha$ equivalent widths are 52.8, 70.3 and $54.1\rm\AA$ for the nucleus, SW disk and inner-N hotspot, with the highest of all, $106.7\rm\AA$ for the far-N hotspot, but the nucleus produces by far the greatest $\rm H\alpha$ luminosity.

   \begin{figure}
 \includegraphics[width=0.86\hsize,angle=-90]{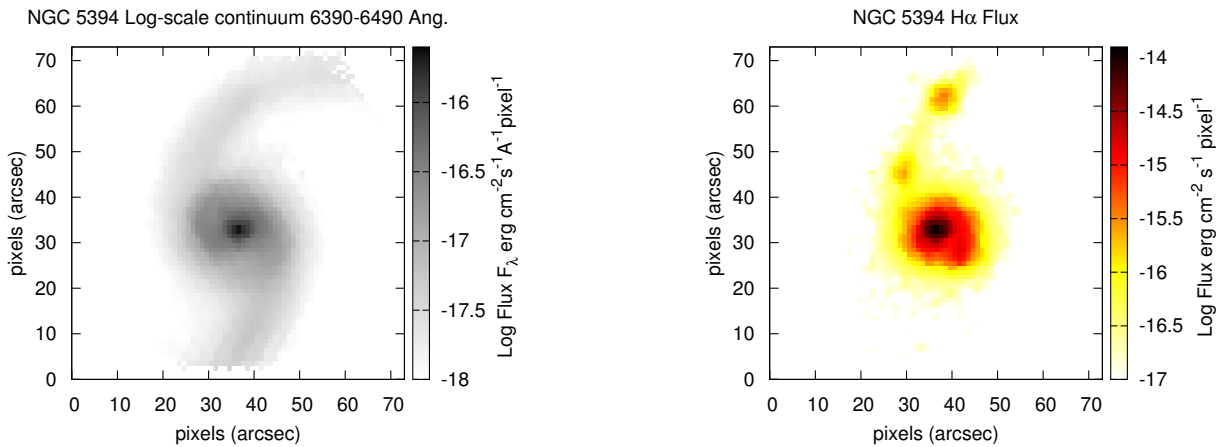}
\caption{NGC5394 red-band continuum map from CALIFA data. North is up and East is left, as in all figures.} 
 \end{figure}

 Arp (1969) described an `unusually sharply defined string of condensations coming out of the companion on the inside of the northern arm'. This `string' is visible on the GTC image, and the inner-N and far-N $\rm H\alpha$-emitting hotspots can be identified with the knots at either end. Probably the intermediate knots are recent sites of star-formation, which now continues only at the extremities. There is  lower intensity $\rm H\alpha$ emission distributed over the whole disk (out to $r\simeq 12$ arcsec), but almost none from the tidal arms outside of the two hotspots. 
 
\begin{figure}
 \includegraphics[width=0.86\hsize,angle=-90]{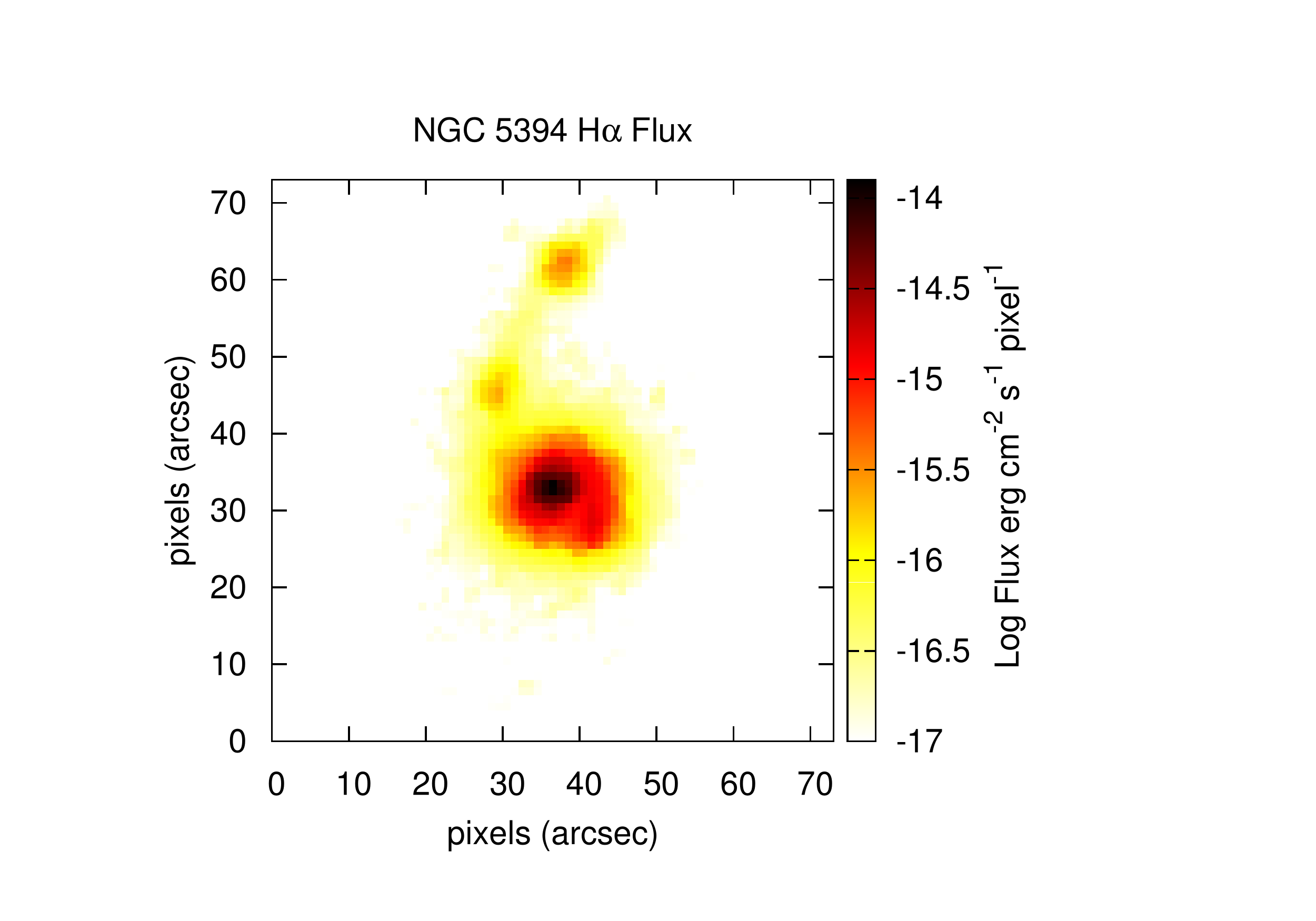}
 \includegraphics[width=0.86\hsize,angle=-90]{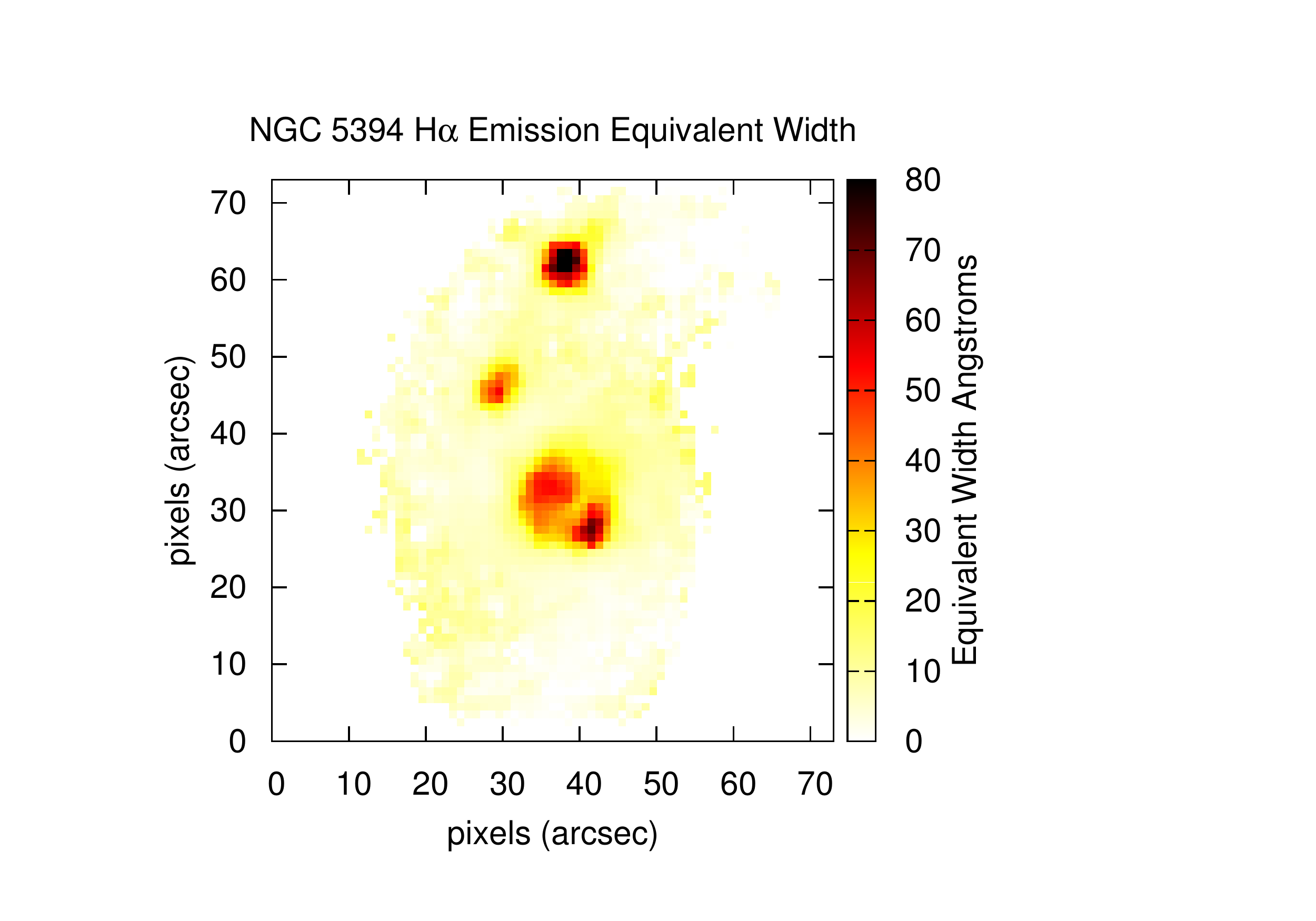}
\caption{Map of (log scale) $\rm H\alpha$ emission-line flux $\rm arcsec^{-2}$ (above) and equivalent width in $\rm \AA$ (below).} 
 \end{figure}

    \begin{figure}
 \includegraphics[width=0.72\hsize,angle=-90]{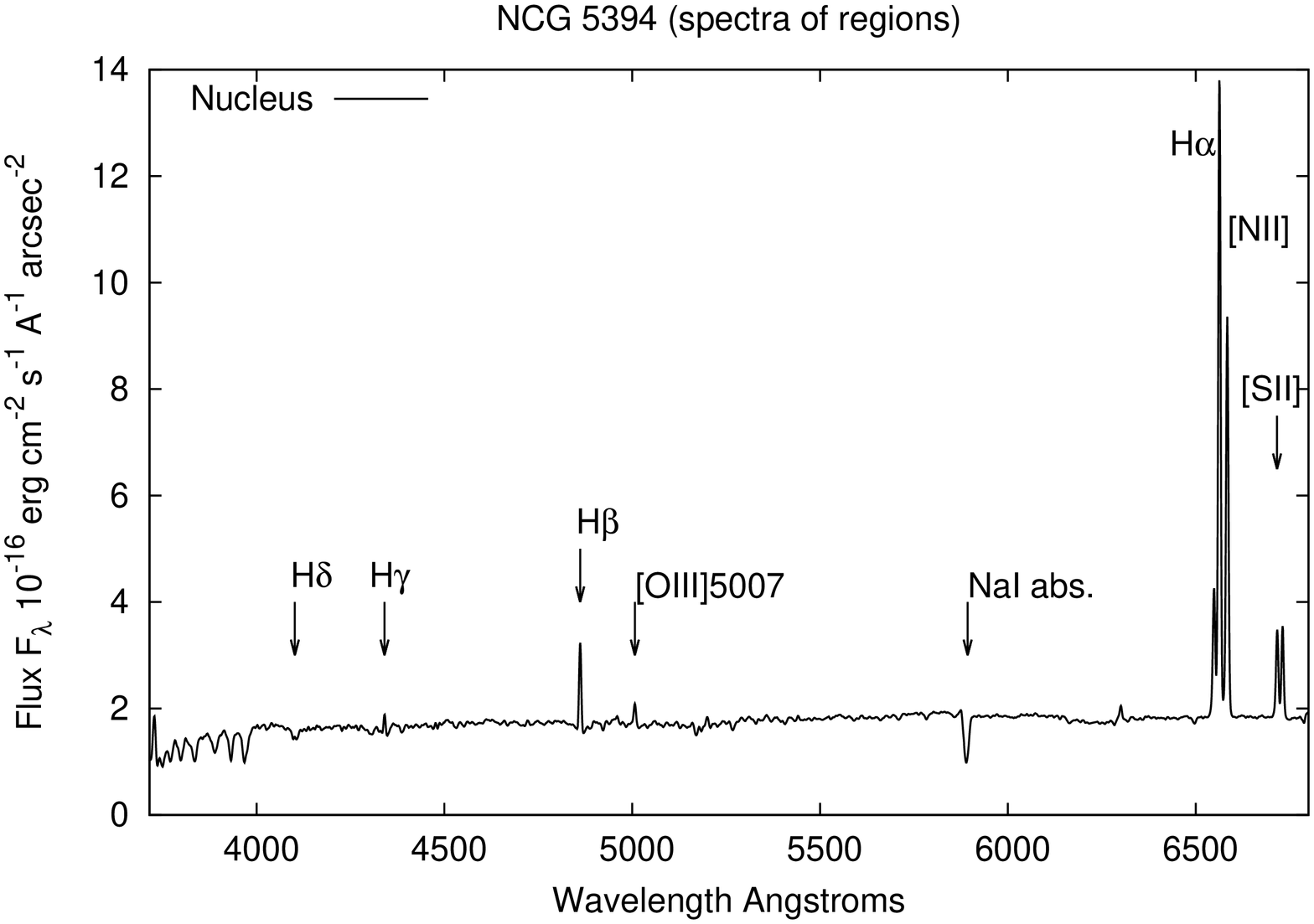}
 \includegraphics[width=0.72\hsize,angle=-90]{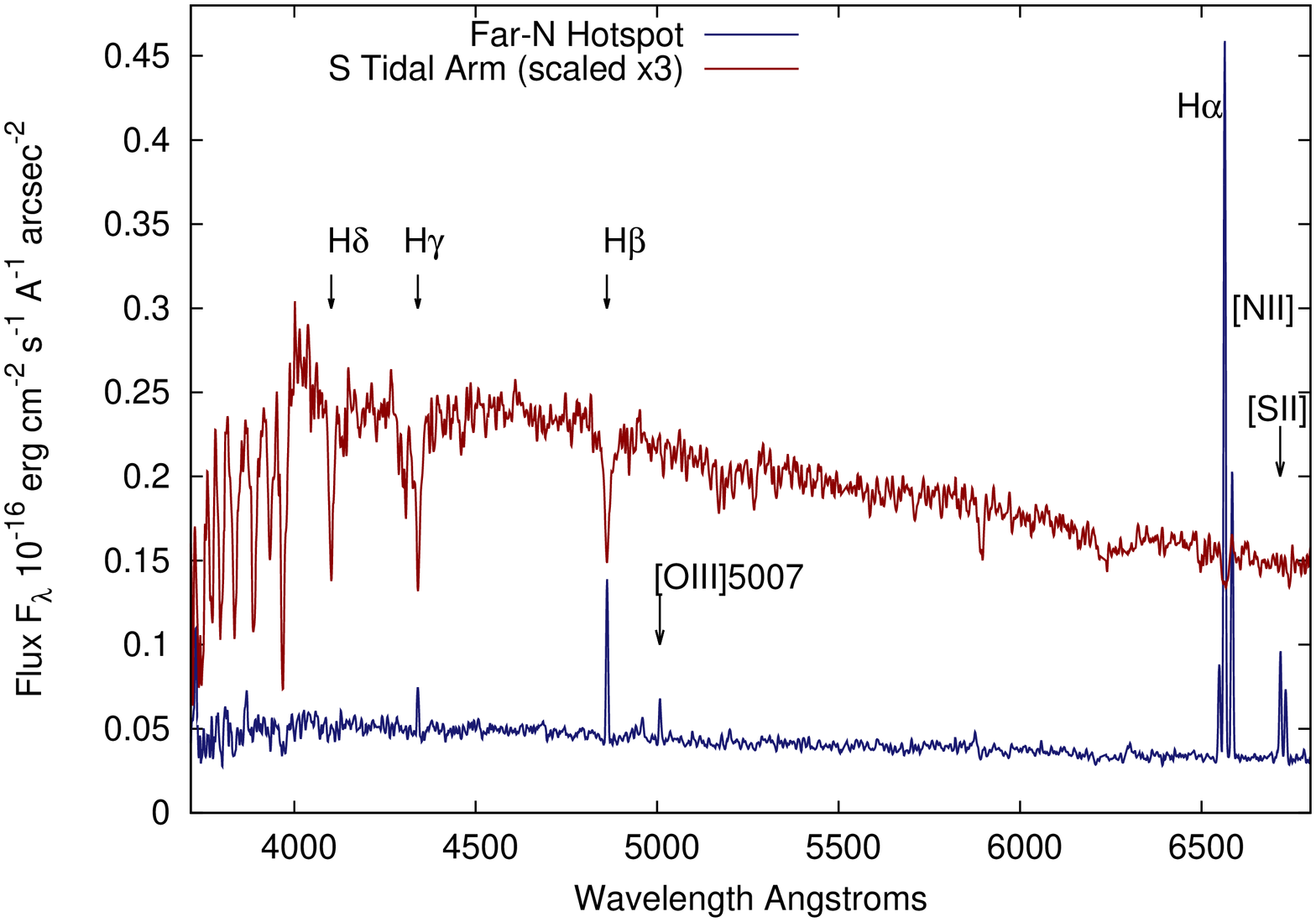}
\caption{CALIFA V500-grating spectra in 3 different regions of NGC 5394, the nucleus, the far-north star-forming hotspot (averaged over central 5 pixels), and the (post-starburst) southern tidal arm. } 
 \end{figure}
 
  To illustrate some of the diversity within this galaxy, Fig 4 compares small-aperture CALIFA spectra of the nucleus, the far-N hotspot (averaged over the central 5 pixels of each) and a region in the southern arm (averaged over 14 pixels). Wavelengths 5508--$5522\rm \AA$, 6222--$6236\rm \AA$ and 6285--$6295\rm \AA$ (galaxy restframe) are severely affected by strong sky OI lines (observed at 5577.3, 6300.3 and 6363.8$\rm \AA$), and are masked out in the plots and model fitting. The nucleus has the strong emission lines expected for a starburst, although [OIII]5007 is relatively weak and the continuum relatively red (i.e. redder than flat in $F_{\lambda}$). The two outer regions are bluer and the N hotspot has strong emission lines, while the S arm is not only passive but may be post-starburst on the basis of strong Balmer absorption lines ($\rm EW>5\rm \AA$). The nucleus has notable differences in emission-line ratios compared to the hotspot, with  higher [NII]6584/$\rm  H\alpha$  and lower [SII]6717/[SII]6731.

 SFRs locally and for the whole galaxy can be estimated by summing the $\rm H\alpha$ fluxes, but there is evidence (K02) the emission lines are substantially attenuated by internal dust. For star-forming HII regions  the intrinsic flux ratio of   $\rm H\alpha/H\beta$ (when corrected for stellar absorption as here) is  expected to be 2.86 (from `case B' models with $\rm T_e\sim10^4K$), but the observed ratio is increased in proportion to the  dust-reddening.  Using the extinction law of Calzetti et al. (2000), supported by the more recent results of Wild et al. (2011), a dust-corrected  $\rm H\alpha$ flux is estimated for each individual pixel  as $\rm F(H\alpha)_{corr}\simeq F(H\alpha)_{obs}[H\alpha/2.86 H\beta]^{2.615}$.   Here this dust correction is only applied for pixels where the statistical error $ \rm  \sigma(H\alpha/H\beta)<1.0$,  which includes all of the disk and the star-forming regions in the northern arm (the other pixels are included in the $\rm H\alpha$ flux summations but uncorrected). Figure 5 shows a map of the $\rm H\alpha$ dust extinction, which is greatest for the nuclear region (2.32 mag).
\begin{figure}
 \includegraphics[width=0.86\hsize,angle=-90]{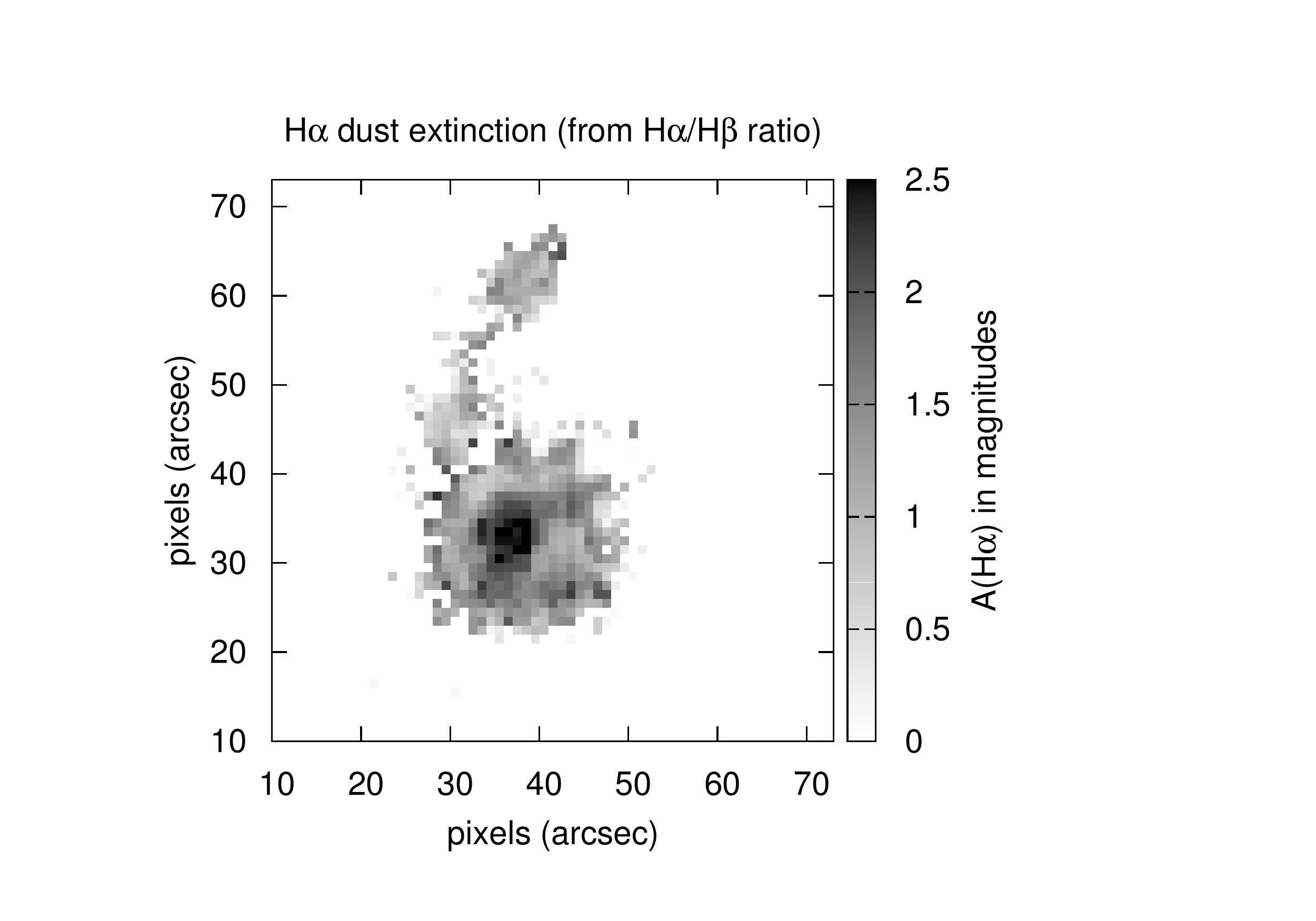}
\caption{$\rm H\alpha$ dust extinction as estimated from the Balmer decrement $\rm H\alpha/H\beta$ using the Calzetti et al. (2000) law.} 
 \end{figure}
 
 To depict the IFS data (without defeating the object by too much spatial binning) can be challenging and three methods are used here, 2D maps, radial profiles averaged in circular annuli, and we also divide the galaxy into 7 regions, shown on Fig 6.
 \begin{figure}
 \includegraphics[width=0.99\hsize,angle=0]{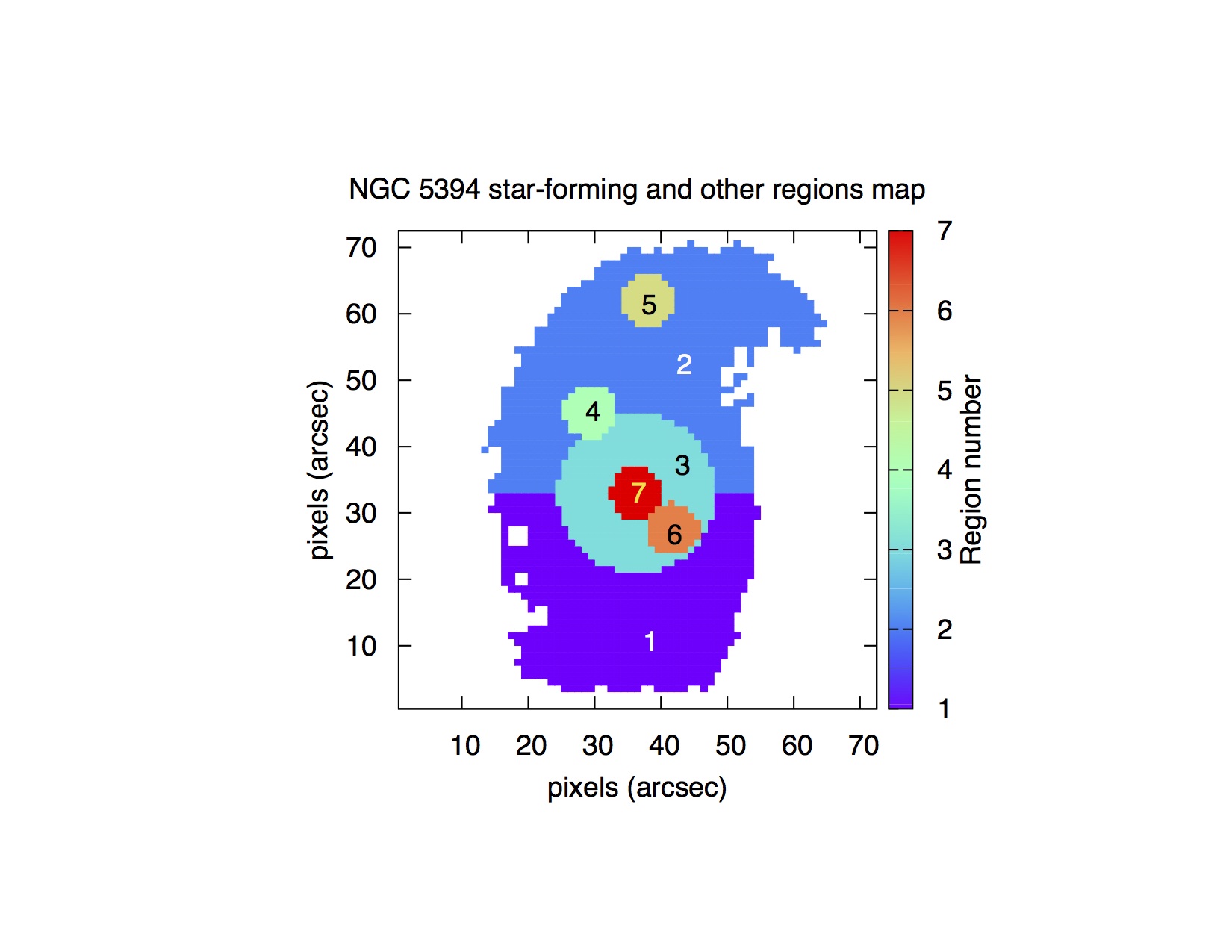}
\caption{Map of regions the galaxy is divided into for the analysis in Table 1: (1) S Arm (2) N arm (3) Outer disk (4) Inner-N hotspot (5) Far-N hotspot (6) SW disk (7) Nucleus} 
 \end{figure}
 Table 1 gives the observed and dust-corrected $\rm H\alpha$ fluxes, summed over each of the 7 regions and for the whole galaxy (for the integrated $\rm H\alpha$ the estimated extinction is 1.98 magnitudes). Fig 7 shows the integrated  CALIFA spectra of each region (note they have different areas), which will be model-fitted in a later Section.
  
\begin{table*}
\begin{tabular}{lccccccccc}
\hline
Region & $\rm F(H\alpha)_{obs}$ & $\rm F(H\alpha)_{corr}$ & $\rm L(H\alpha)_{corr}$ & $\rm SFR_{corr}$ & $\rm [NII]\over H\alpha $ & $\rm [OIII]\over H\beta$ &  $\rm [SII]\over H\alpha$ & $\rm [OI]\over H\alpha$  & [SII]$6717\over 6731$ \\
  & erg $\rm cm^{-2}s^{-1}$ & erg $\rm cm^{-2}s^{-1}$ & erg $\rm s^{-1}$ & $\rm M_{\odot}yr^{-1}$ & &  & &  &\\
    \hline
Nucleus   &   $1.981\times10^{-13}$ & $1.681\times 10^{-12}$ &  $5.749\times 10^{41}$ & 2.53 & 0.592 & 0.273 & 0.288 & 0.023 & 1.069  \\
SW disk  & $3.776\times 10^{-14}$  & $1.547\times 10^{-13}$ & $5.290\times10^{40}$ & 0.23 & 0.452  & 0.262  & 0.296 & 0.031 &  1.375 \\
Outer disk & $9.414\times 10^{-14}$ & $3.603\times 10^{-13}$ & $1.232\times 10^{41}$ & 0.54 & 0.551 &  0.407 & 0.343 & 0.043 &  1.351 \\
Inner-N hotspot & $5.273\times 10^{-15}$ & $1.154\times 10^{-14}$ & $ 3.946\times 10^{39}$ & 0.017 & 0.466  & 0.425 & 0.289 & 0.033 & 1.301  \\
Far-N hotspot & $7.257\times 10^{-15}$  & $2.031\times 10^{-14}$ & $6.946\times 10^{39}$ & 0.031 
& 0.408 & 0.280 & 0.270 & 0.019 & 1.526 \\
N arm   & $1.161\times 10^{-14}$ & $1.578\times 10^{-14}$ & $5.396\times 10^{39}$ & 0.024 &  &  & & & \\
S arm   & $8.861\times 10^{-15}$ & $1.088 \times 10^{-14}$ & $3.721\times 10^{39}$ & 0.016 & & &   & & \\
 Whole Galaxy    & $3.629\times 10^{-13}$ & $2.242\times 10^{-12}$ & $7.710\times 10^{41}$ & 3.39
 & 0.562 & 0.368 & 0.312 & 0.033 & 1.191 \\
\hline
\end{tabular}
\caption{Integrated $\rm H\alpha$ luminosities for each of 7 regions of the galaxy NGC 5394 (Fig 6); fluxes and luminosities corrected for dust using the Balmer decrement; star-formation rates derived from the corrected $\rm L(H\alpha)$ (Chabrier IMF); and several important line ratios from integrated, uncorrected fluxes (not shown for tidal arms as the line fluxes are so low).}
\end{table*}
  
      \begin{figure}
 \includegraphics[width=0.72\hsize,angle=-90]{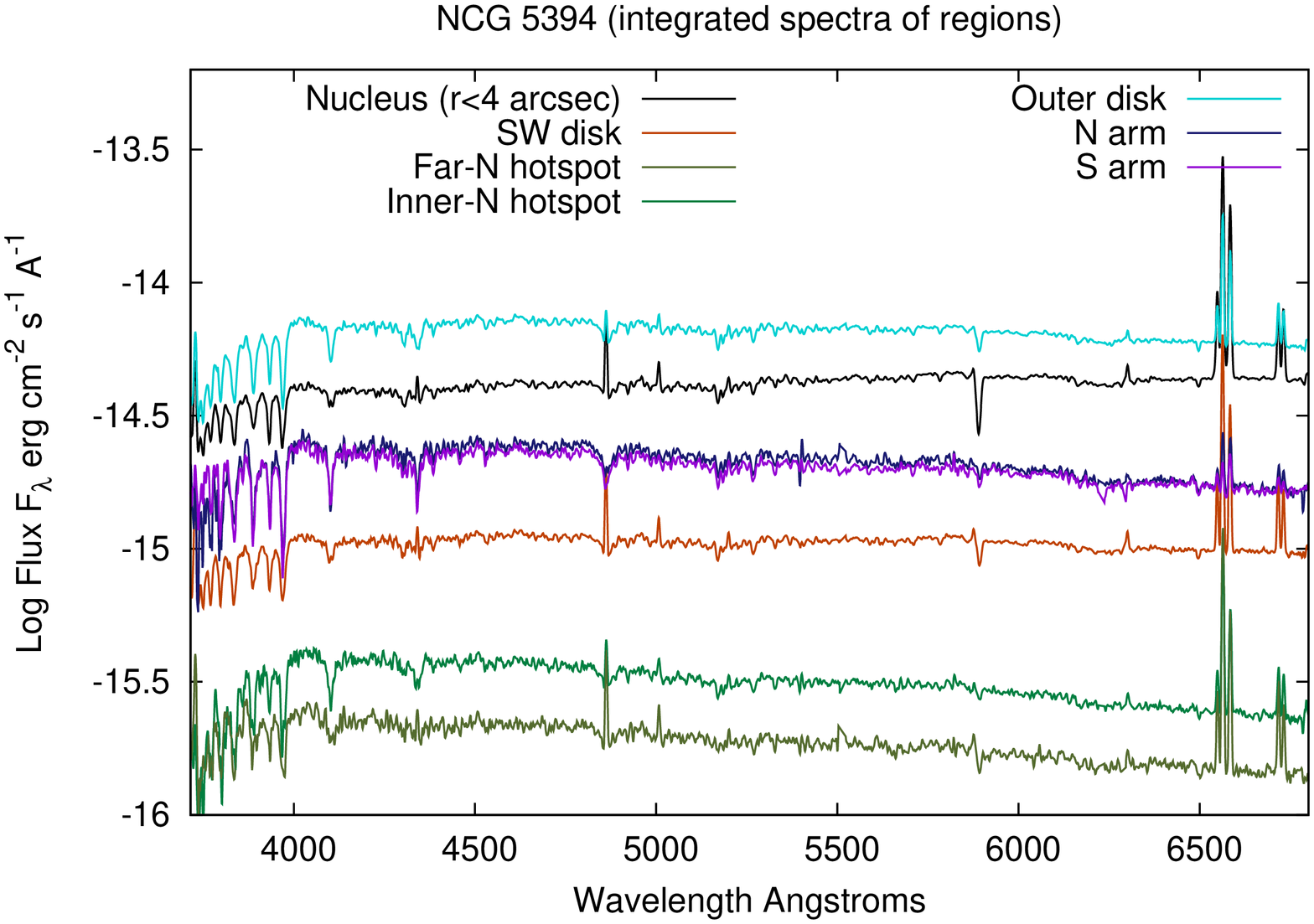}
\caption{CALIFA V500-grating spectra integrated over 7 regions of NGC 5394 as shown in Fig 6, on a log scale. } 
 \end{figure}
  
   From the dust-corrected fluxes, $\rm H\alpha$ luminosities in erg $\rm s^{-1}$ are calculated using the distance modulus 33.64 mag. A star-formation rate in solar masses ($\rm M_{\odot}$) per year can then be estimated as $\rm 4.4\times10^{-42}L(H\alpha)$
(e.g. Sobral et al. 2014); this conversion factor is for the Chabrier IMF. We estimate  $3.39~ \rm M_{\odot} yr^{-1}$ for the whole galaxy. For the Salpeter IMF (with more low mass stars) the SFR would be $7.9\times10^{-42}\rm L(H\alpha)$ (Kennicutt 1998) giving $6.09 ~\rm M_{\odot} yr^{-1}$, in agreement with the K02 estimate $6\pm \rm 2 M_{\odot} yr^{-1}$.  In comparison, for the Mice galaxies (each larger and about twice as massive as NGC 5394), Wild et al. (2014) estimated SFRs $6.2 \rm ~M_{\odot} yr^{-1}$ for NGC 4676A and $2.3  \rm ~M_{\odot} yr^{-1}$ for NGC 4676B  (correcting for dust using UV and mid-infrared fluxes). 

We estimate the far-N hotspot accounts for 0.9\% of the galaxy's SFR, the same as its  fraction of the  $8\mu m$ flux (2.626/279.8 mJy) in {\it Spitzer} observations (Higdon et al. 2014). Such features are not unexpected e.g. the `ocular' IC 2163 has at least 9 star-forming clumps spaced around the `eye', with similar $\rm H\alpha$ and $8\mu m$ fluxes (Kaufman et al. 2012). In NGC 5394, star-formation is now concentrated in the nucleus, with 75\% at $\rm r<1$ kpc.
 \begin{figure}
 \includegraphics[width=0.72\hsize,angle=-90]{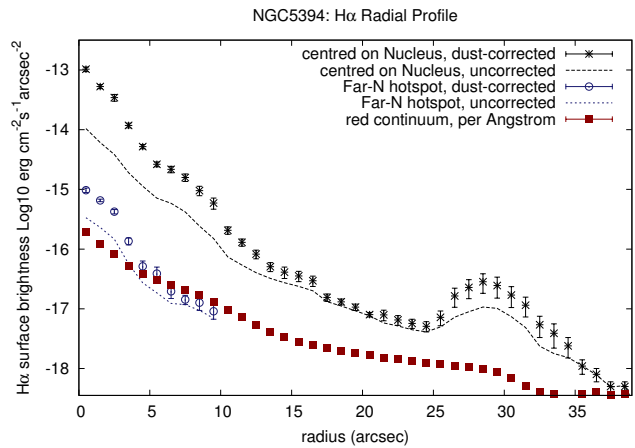}
\caption{Radial profile of $\rm H\alpha$ emission, centred on the nucleus, as observed and as corrected for dust using the $\rm H\alpha/H\beta$ ratio. Also shown, at $r<10$ arcsec, the profile of the hotspot in the northern arm, and for comparison, the shallower profile of the red continuum (6390--$6490\rm\AA$) in flux $\rm \AA^{-1}$.} 
 \end{figure}
  Fig 8 shows the observed and dust-corrected $\rm H\alpha$ flux profile, averaged in circular annuli centred on the nucleus (the peak at $r=29$ arcsec is the far-N hotspot), and also profiles centred on the far-N hotspot itself out to $r=10$ arcsec only. The nucleus and hotspot are very different in luminosity but are concentrated star-formation regions and have similarly steep profiles. In the continuum, the profile of the galaxy is shallower, closer to an exponential, without the steepening at $r\leq 10$ arcsec.
  The central star-formation density is very high (higher than in NGC 4676A), with an integrated $\rm H\alpha $ flux  $1.03\times 10^{-13}$ (observed) or $9.39\times 10^{-13}$ (corrected) in the central 14 pixels (0.889 $\rm kpc^2$); the latter corresponds to  1.6 $\rm M_{\odot}yr^{-1}kpc^{-2}$. 
  
 \section{Line ratios}
 We examine a number of other diagnostic lines and ratios. First, the equivalent width of the $\rm H\delta$ stellar absorption line (Fig 9) is high (6--$9\rm\AA$) over most of the galaxy area, especially the tidal arms. An $\rm  EW_{abs}( H\delta)>5 \AA$ is considered diagnostic of a post-starburst region or galaxy (e.g. Goto 2008) and is caused by an enhancement in the fraction of A-type stars, which on the basis of stellar lifetimes implies they formed in  a period of increased star-formation between 100 Myr and 1 Gyr ago. Only in the active star-forming ($\rm H\alpha$-emitting) regions is $\rm EW(H\delta)$ reduced to a more typical 3--$4\rm \AA$, presumably because the A-type spectrum is mixed with even younger OB stars. This shows NGC 5394 is very much a post-starburst galaxy with the earlier burst more spatially extended than the current activity.
 \begin{figure}
 \includegraphics[width=0.86\hsize,angle=-90]{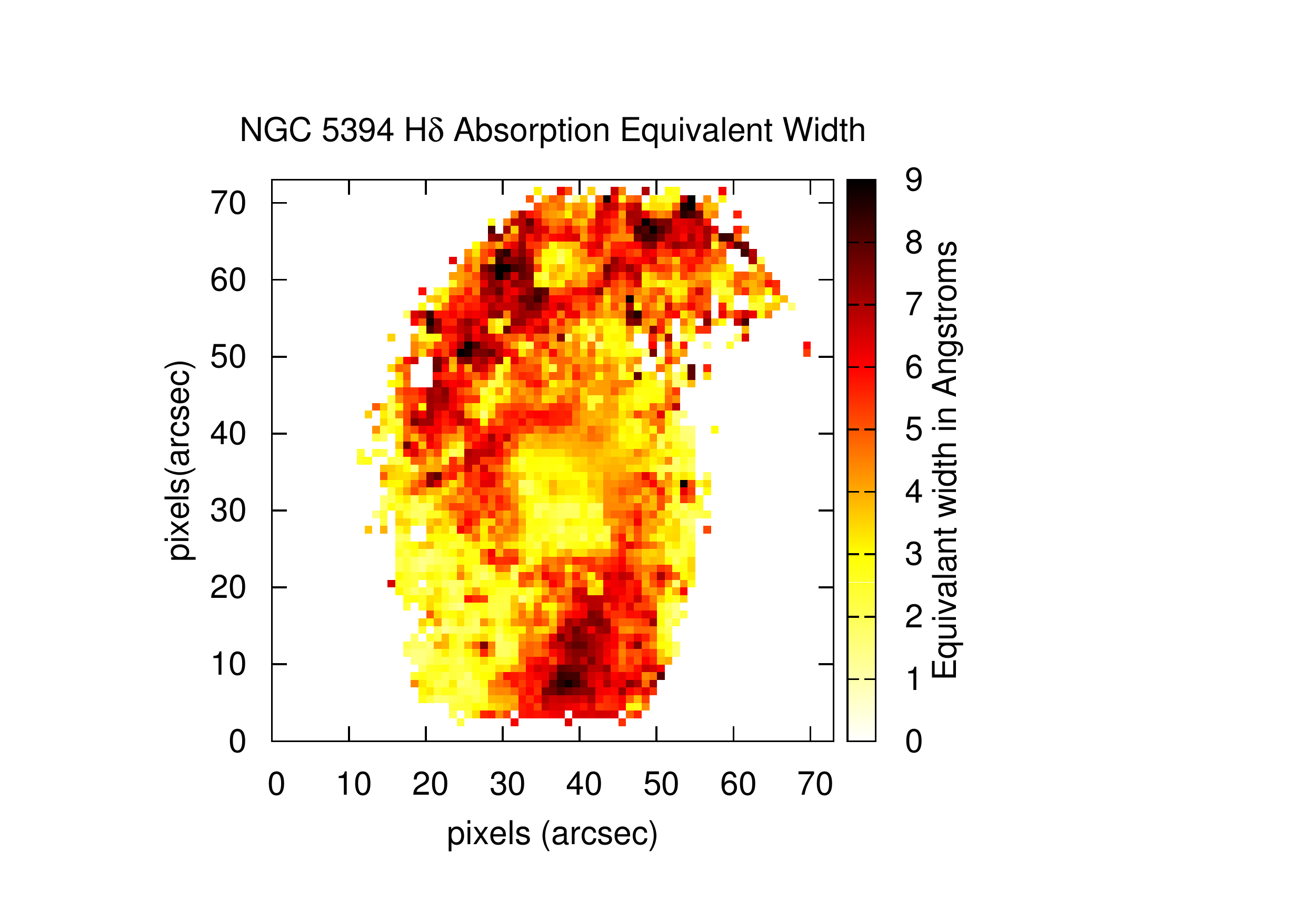}
\caption{Map of $\rm H\delta$ absorption-line equivalent width.} 
 \end{figure}
 
 Of much importance in identifying AGN and other sources of high excitation, and estimating metallicity, are the line ratios $\rm [NII]6584/H\alpha$ (the N2 index) and $\rm [OIII]5007/H\beta$ (the O3 index). These and the other line ratios considered here (except  $\rm H\alpha/H\beta$) are almost insensitive to dust because the two wavelengths are very close. 
 For the galaxy as a whole $\rm O3=0.368$ and $\rm N2=0.562$. Fig 10 and 11 show the ratio maps, and 
 Fig 12 the radial profiles (centred on the nucleus, and for comparison at the far-N hotspot). 
 Many pixels, generally in the outer galaxy where emission lines are weak or absent, give very noisy or meaningless ratios, and
   these maps and profiles include only  pixels where the ratio measurement is $>3\sigma$.  
 \begin{figure}
 \includegraphics[width=0.86\hsize,angle=-90]{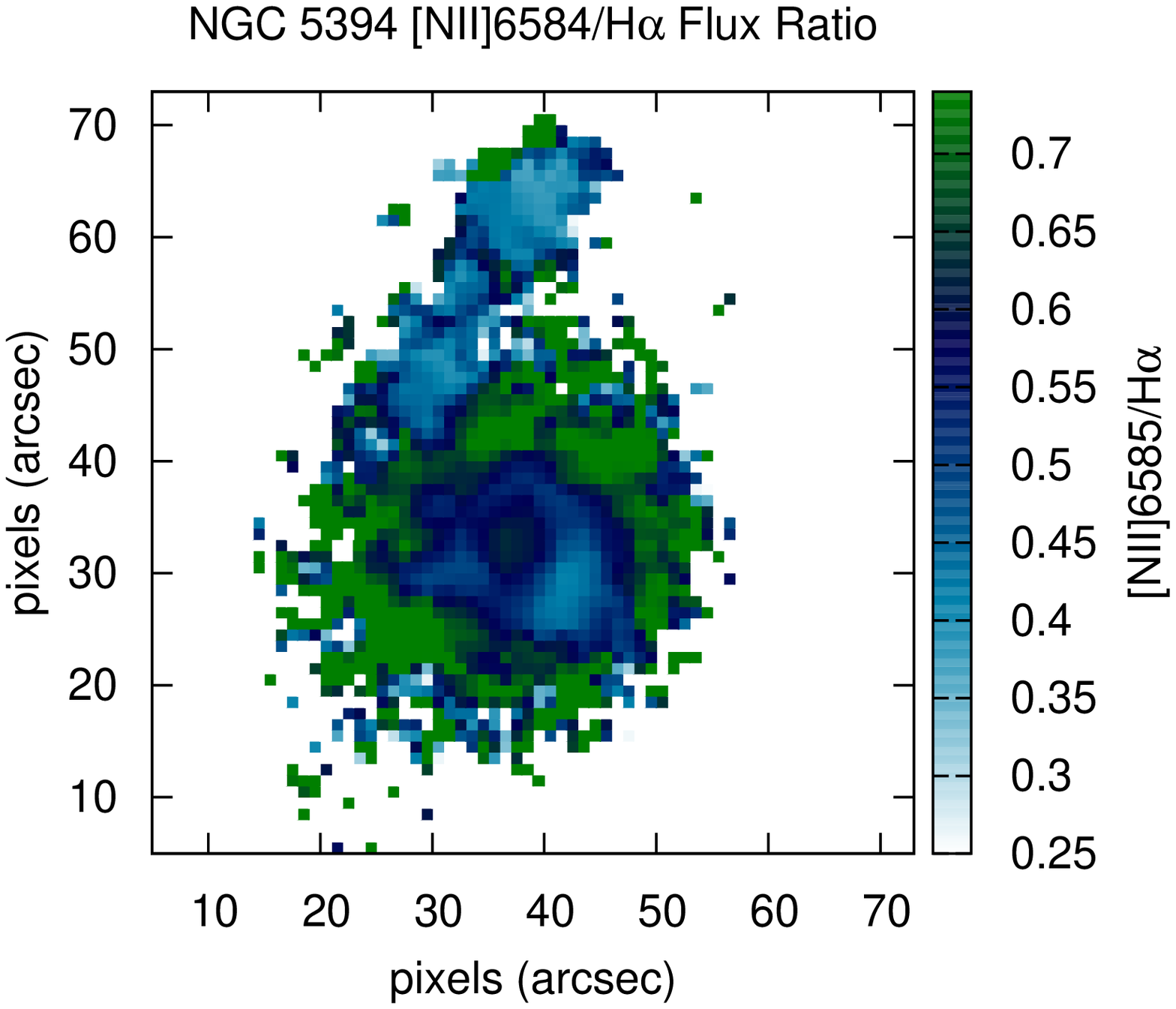}
\caption{Map of $\rm [NII]/H\alpha$ ratio (N2 index).} 
% \end{figure}
 %\begin{figure}
 \includegraphics[width=0.86\hsize,angle=-90]{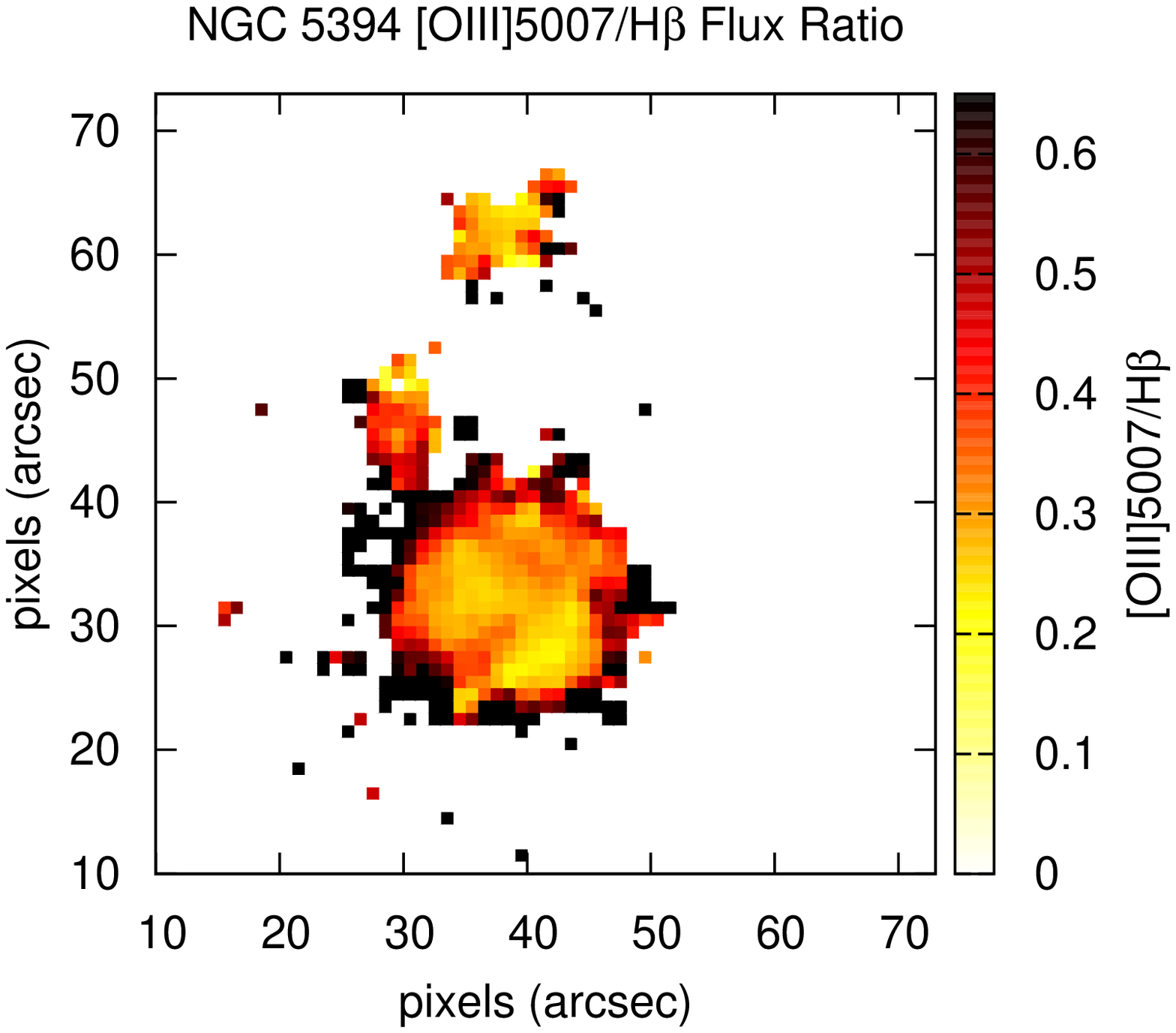}
\caption{Map of $\rm O[III]5007/H\beta$ ratio (O3 index).} 
 %\end{figure}
%\begin{figure}
 \includegraphics[width=0.72\hsize,angle=-90]{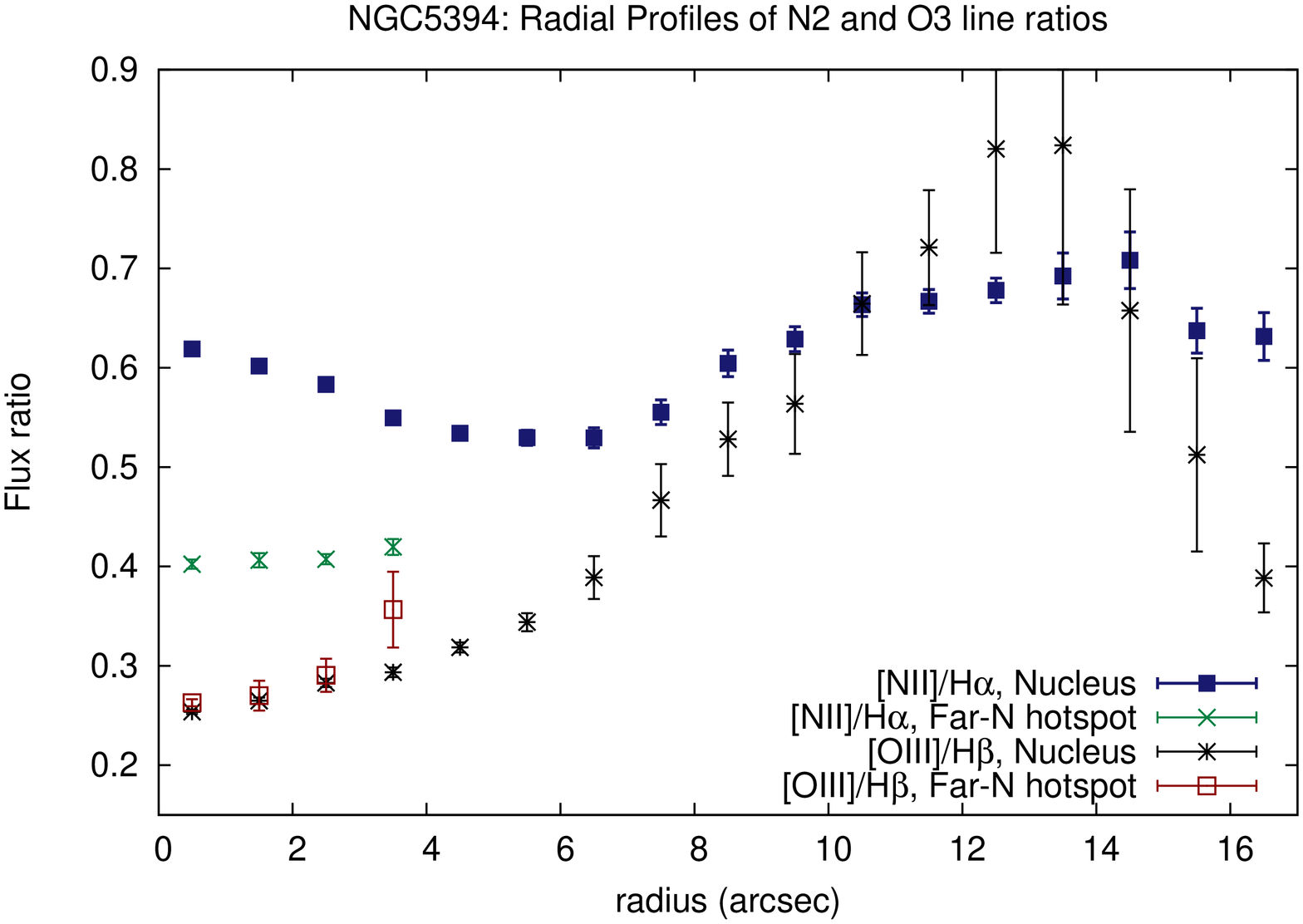}
\caption{Radial profiles of the line ratios N2 and O3 (centred on the nucleus), and for comparison the profiles centred on the far-N hotspot.} 
 \end{figure}
 
 The O3 ratio is low at 0.25--0.3 within the main star-forming regions, with no increase at the centre (only 0.254 for the central 4 pixels), so gives no evidence for an AGN. In contrast, N2 is strongly peaked at the nucleus, 0.619 for the central 4 pixels and 0.629 at the highest pixel, but falls to 0.52 at $r=6$ arcsec, and 0.41 in the far-N hotspot, which is more typical for star-forming regions. However, at the outer edge of the disk the mean N2 increases again to $>0.6$, and   in many pixels here it is even higher than at the centre, with ratios 0.65--0.85. At the same radii there is an increase in O3, and to some extent in $\rm [SII]/H\alpha$ and $\rm [OI]/H\alpha$. We find an  outer  region of enhanced line ratios forming an irregular  annulus, mostly  between $r\simeq 9$ and 16 arcsec.
The N2 ratios in both the nucleus and this outer region exceed the log (N2)=-0.25 upper limit of typical star-formation (e.g. the Kauffmann et al. 2003 divide).

 \begin{figure}
 \includegraphics[width=0.86\hsize,angle=-90]{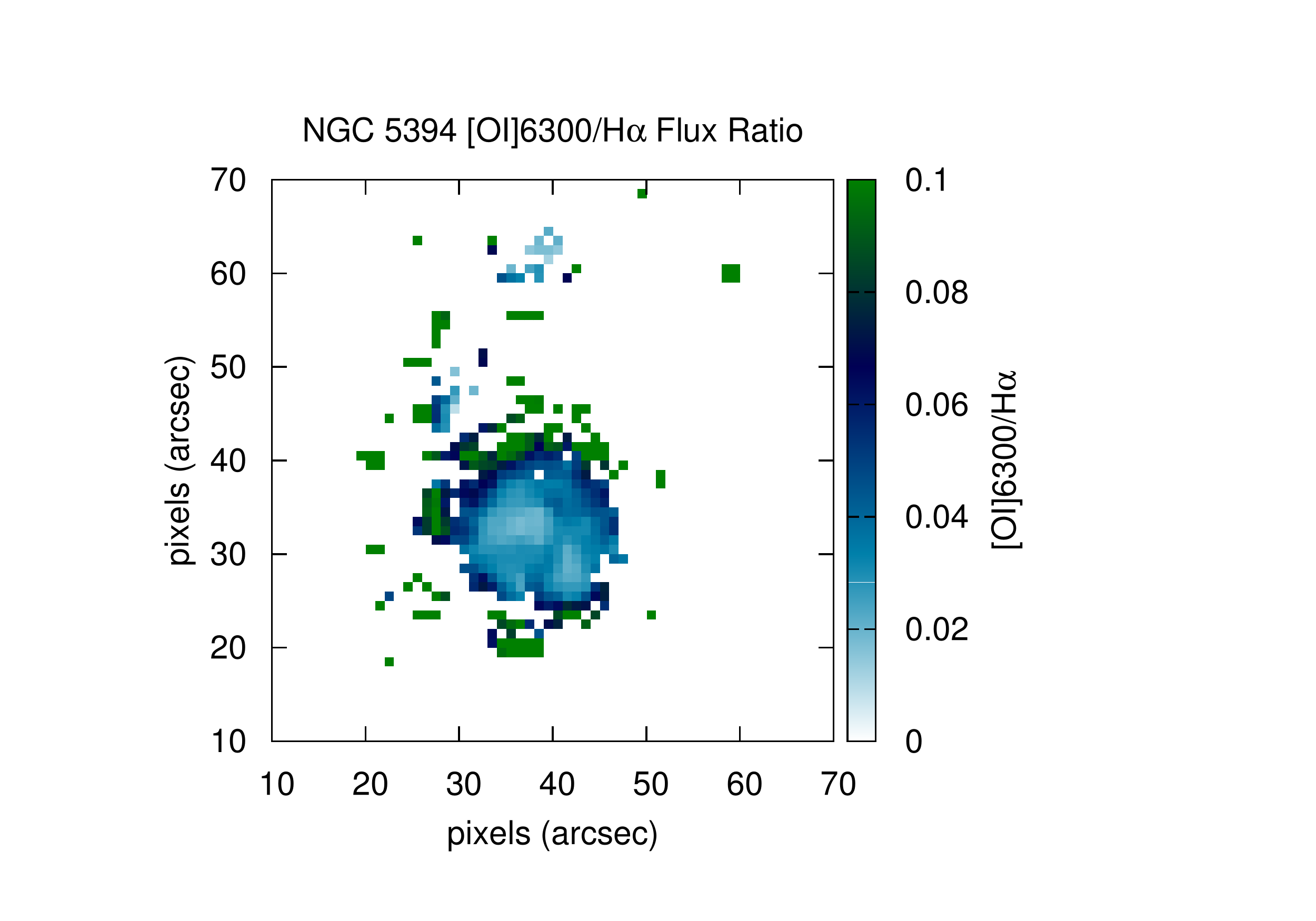}
\caption{Map of $\rm [OI]6300/H\alpha$ ratio.} 
 \end{figure}   
    
    \begin{figure}
 \includegraphics[width=0.86\hsize,angle=-90]{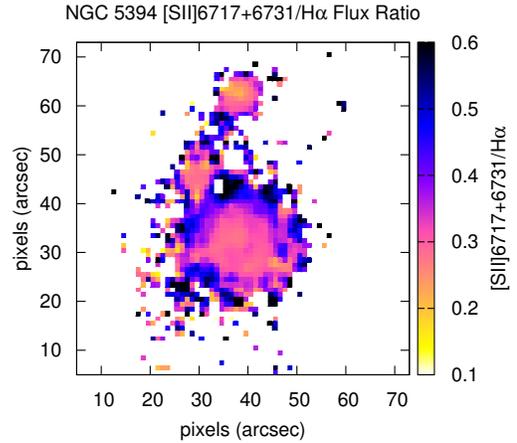}
\caption{Map of the S[II]6717+6731$\rm/H\alpha$ ratio.} 
 \end{figure}

   \begin{figure}
 \includegraphics[width=0.72\hsize,angle=-90]{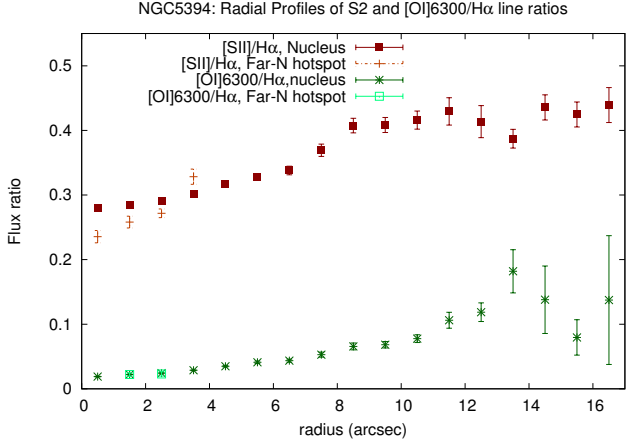}
\caption{Radial profiles of the line ratios $\rm OI/H\alpha$ and $\rm [SII](6717+6731)/H\alpha$ in the disk, and centred on the far-N hotspot.} 
 \end{figure}
 
 \begin{figure}
 \includegraphics[width=0.86\hsize,angle=-90]{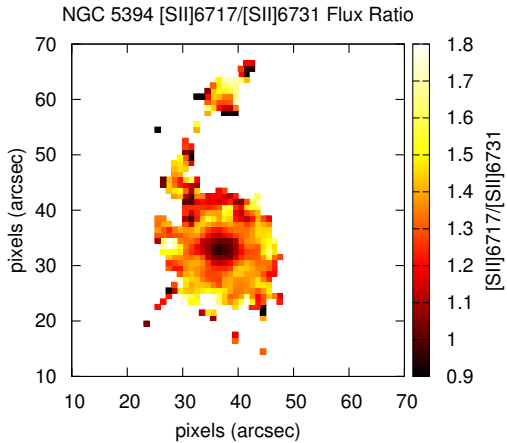}
\caption{Map of [SII]6717/[SII]6731 ratio, sensitive to electron density $n_e$} 
 \end{figure}

  \begin{figure}
 \includegraphics[width=0.72\hsize,angle=-90]{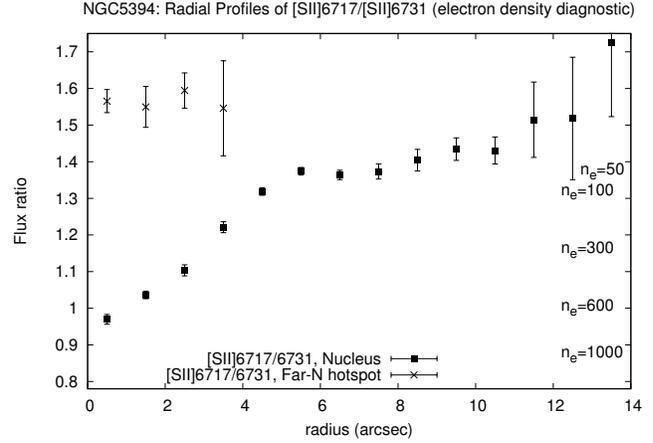}
\caption{Radial profiles of the line ratios $\rm [SII]6717/[SII]6731$ in the disk, and centred on the far-N hotspot. An $n_e$ scale is shown for $T_e=10000K$.} 
 \end{figure}
   
   CALIFA data for the Mice galaxies (Wild et al. 2014) showed large biconical regions in both disks with $\rm N2>-0.25$ dex (0.56) up to $\rm N2\simeq1$, with spatially matching high values for the ratios $\rm [OI]6300/H\alpha$ (with 0.1-0.25) and $\rm [SII]/H\alpha$ (with 0.6-1.0), and (in 4676A) O3 (which ranged from 1--4). These high ratios and the bicone morphology could be explained by high-velocity ($\sim 350$ km $\rm s^{-1}$) shocks with `ionized precursor',  from a super wind, an outflow driven by a central starburst.   
  If the nuclear starburst in NGC 5394 causes similar shock ionization, this would appear directly in front of the nucleus due to the face-on orientation. However, at the nucleus, O3 and the  other two key ratios $\rm [OI]6300/H\alpha$ and $\rm [SII](6717+6731)/H\alpha$ (Fig 12 to 15) do not show high values or any peak relative to larger radii, and all three  (in contrast to the  N2 ratio), remain within the range expected for HII regions, and so give no evidence of shocks (or AGN) here. 
  
 The flux ratio of the two lines in the [SII] 6717,6731 doublet is a useful diagnostic of electron density ($n_e$) in nebulae, most sensitive at $n_e\sim 10^2$--$10^4\rm cm^{-3}$ (with also a small dependence on temperature, see e.g. Cant\'o et al. 1980). In NGC 5394, the [SII]6717/6731 ratio (Fig 16 and 17) shows a strong monotonic gradient at $r<6$ arcsec, falling from 1.4 in the outer disk ($r\sim 6$--12 arcsec) to 0.97 at the central 5 and 0.949 in the central one pixel. For the calibration in {\it iraf.stsdas.nebula.temden}, assuming $\rm T_e=10000K$, this signifies a rise in electron density from $n_e\simeq 27$ $\rm cm^{-3}$ to a central 750 $\rm cm^{-3}$. The higher  [SII]6717/6731 ratio of 1.55-1.6 in the far-N hotspot indicates a lower density, probably $n_e\leq  20$ $\rm cm^{-3}$.  For the galaxy as a whole the integrated [SII]6717/6731 ratio is 1.191 (corresponding to 252 $\rm cm^{-3}$).  
 
In NGC 4676,  Wild et al. (2014) found intermediate [SII]6717/6731 ratios of 1.1--1.2 ($n_e\simeq 200$--400 $\rm cm^{-3}$) in the bicones and both disks, without  obvious radial gradients. The non-interacting M33 spiral, studied with the same PMAS spectrograph (Lopez-Hernandez et al. 2013), does show a gradient from an outlying HII region with $\rm [SII]6717/6731\simeq 1.5$--1.8 (very low $n_e$), to the nuclear HII region with 1.30 ($n_e\sim  100$), to the very centre  with 1.07 ($n_e\simeq 328~\rm cm^{-3}$). In the starburst-merger Arp 299 (Heckman et al. 1999)  $n_e$ similarly increases from $\sim 50$ to 250--300 $\rm cm^{-3}$ at the centre.  Le Tiran et al. (2011) found the stacked spectra of $z=1.3$--2.6 starburst galaxies to indicate mean $n_e$ values of 200 $\rm cm^{-3}$ (range 100--500) for the $\rm H\alpha$-brightest regions, similar to local starbursts. On this basis, the $n_e\sim 750$ $\rm cm^{-3}$ in the NGC 5394 nucleus is a fairly extreme case, probably a consequence of the interaction.   
  \begin{figure}
 \includegraphics[width=0.72\hsize,angle=-90]{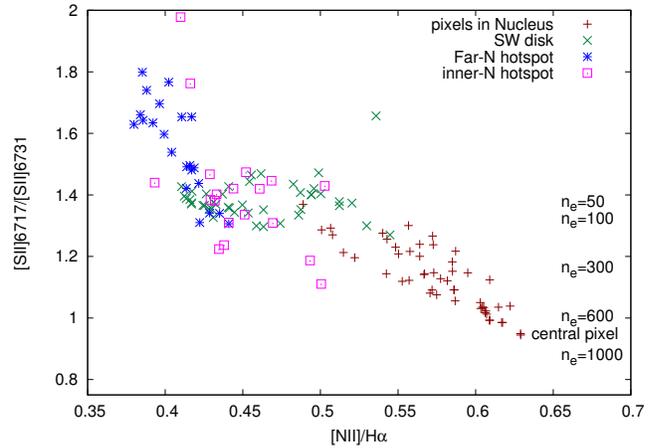}
\caption{Plot of the line ratios N2 vs.  $\rm [SII]6717/[SII]6731$ in the strongly star-forming regions, showing that they are correlated (the correlation coefficient is 0.8375) forming a sequence of increasing $n_e$. An $n_e$ scale is shown for $T_e=10000K$.} 
 \end{figure}
 
 Kewley et al. (2013) predict in stellar photoionization models that   
 increasing $n_e$ from 10 to 1000 $\rm cm^{-3}$ could, by increasing collisional excitation, raise the N2 ratio by as much as 0.2 dex (their Fig 2; note that O3 is also increased by a similar amount). This might in itself account for the central high peak of N2 in NGC 5394, even without shocks or AGN. Observationally a high $n_e$ is associated with a positive N2 offset in the BPT diagram for  SDSS galaxies (Brinchmann, Pettini \& Charlot 2008) and for individual HII regions in CALIFA (Fig 2 of S\'anchez et al. 2015).
  Plotting for the individual pixels in the 4 main star-forming regions (Fig 18), N2 is closely correlated with the [SII]6717/6731 ratio, which supports  $n_e$ as the primary cause of the central enhancement.
  
   In contrast, the outer, annular region of high N2 and other ratios is not associated with any increase in $n_e$. To confirm this region is genuine and investigate further we sum the spectra of all 296 pixels at $9<r<16$ arcsec with $\rm N2>0.65$ to give the spectrum shown in Fig 19. $\rm H\alpha$ and [NII]6584 are well-detected, with a ratio that appears close to unity on the plot, but in the integrated line fluxes from {\sevensize PORTO3D} $\rm H\alpha$ is corrected upward for stellar absorption and the resulting mean N2 for these pixels (weighted by flux rather than area) is 0.741. In the same way the other mean ratios are found to be $\rm [OIII]/H\beta=0.844$, $\rm [SII]/H\alpha=0.432$ and $\rm [OI]/H\alpha=0.060$. 
The [SII]6717/6731 ratio is 1.37, giving a low $n_e$ of 20 $\rm cm^{-3}$.     The $\rm H\alpha$ emission and $\rm H\delta$ absorption EW are both quite moderate at $4.7\rm\AA$ and $4.1\rm \AA$ so this is neither a starburst or a predominantly post-starburst  region. 

The elevated line ratios in this region resemble (but are a little more moderate than) the shock-dominated  outer regions of the late merger NGC 3256 (Rich et al. 2011), the extensive eastern shock region of the colliding NGC 7318 in Stephan's Quintet (Rodr\'iguez-Baras et al. 2014) and the bicones in the Mice. All 4 ratios would fit the models of Rich, Kewley \& Dopita (2011) for composite shock excitation and HII regions, where the shock component is $\sim 50\%$ (of the $\rm H\alpha$ flux). Further comparison  with the models tabulated  by Allen et al. (2008) finds a dominant shock component that can fit the observed line ratios quite well with shock velocity $v\simeq 225$--275 km $\rm s^{-1}$ and little or no pre-ionization (which would give a higher O3); this assumed near solar metallicity, magnetic equipartition ($\rm B=3.23\mu G$) and a low electron density $n_e=1~\rm cm^{-3}$.

  \begin{figure}
 \includegraphics[width=0.72\hsize,angle=-90]{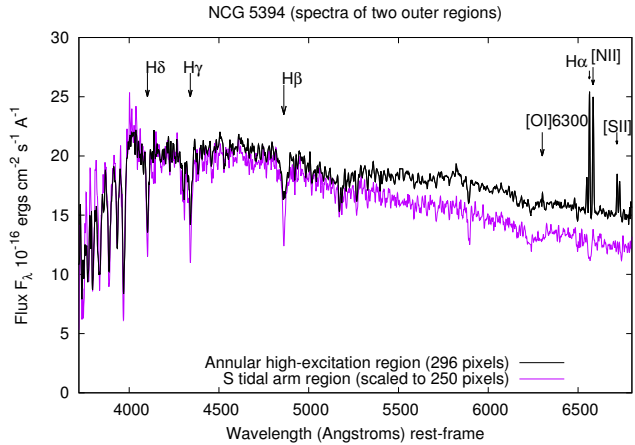}
\caption{Summed spectrum of the annular high-excitation region where the N2 ratio exceeds 0.65, in total  296 $\rm arcsec^2$. Also shown for comparison is the (scaled-up) spectrum from a 14 pixel region of the S tidal tail.}   
 \end{figure}

The total $\rm H\alpha$ flux (uncorrected for dust) from this shock region is   $9.4\times10^{-15}$ erg $\rm cm^{-2}s^{-1}$ ($\rm L=3.2\times 10^{39}$ erg $\rm s^{-1}$)  only $2.6\%$ that of the whole galaxy, so shocks remain a  small (few percent) component of the galaxy's emission-line luminosity compared to star-formation.
Note that this region only appears annular in ratio maps, while all the line fluxes increase towards the centre, so it is possible this emission component extends over the whole disk, but is not apparent inward of $r\sim 4$ kpc where star-formation  dominates the $\rm H\alpha$ flux with a lower N2. In e.g., the elliptical NGC 4696, which has shocks from an ongoing merger but very little star formation (Farage et al. 2010), the N2 ratio remains uniformly close to 2. 

Similar high line ratios have also been observed in some passive, non-interacting galaxies, e.g. a mean N2 ratio 2.2 in the early-type CALIFA sample of Papaderos et al. (2013), which may be the result of hard continua from post-AGB (Asymptotic Giant Branch) stars, planetary nebulae or `hot low mass evolved stars', but in these cases the lines are fainter, with $\rm EW\leq 2\AA$. 
 Fig 19 shows for comparison a scaled-up spectrum from a region in the southern tidal tail (also on Fig 4). The spectra are quite similar except the tail is a little bluer (less dust?), has stronger Balmer absorption, but lacks the emission lines (there is possibly a weak [NII] line but the EW is no more than $0.8\rm \AA$). This is the part of the galaxy most dominated by the intermediate age stellar population. These stars cannot be making more than a very small contribution to the high excitation region, and we look elsewhere for the power source (see section 6.3).

\section{Metallicity gradients and the BPT diagram}
Metallicities of galaxies or sub-regions can be estimated by a great variety of methods. One of the  simplest requires only O3 and N2 ratios to estimate  the nebula gas metallicity, and can be applied even to high redshifts, with dust having little effect, but these ratios are subject to other influences (especially AGN) and there are uncertainties in calibration. The O3N2 index, given as log10[O3/N2], is -0.387 at the centre of NGC 5394 and -0.184 for the galaxy as a whole. In the widely used PP04 (Pettini \& Pagel 2004) calibration,  metallicity expressed as the logarithmic ratio of oxygen to hydrogen, $\rm 12 + log(O/H) = 8.73 - 0.32 \times O3N2 $. Marino et al. (2013), comparing line ratios from the CALIFA survey with calibrations based on electron temperature, derived a shallower relation, $\rm 12 + log(O/H) = 8.533 - 0.214\times O3N2$.

A different metallicity measure is obtained from the stellar component. Stellar and nebular metallicity may differ, as stars preserve a record of the abundances at the time of their formation, which may be Gyr ago, and gas may move in or out. {\sevensize PORTO3D} model fitting generated maps of luminosity-weighted and mass-weighted stellar metallicity (lwZ and mwZ), the former giving more weight to recently formed stars, in units of Z (total mass fraction of $\rm N_{atomic}\geq 3$ elements),  which are here converted to O/H on the basis that solar metallicity is $\rm Z=0.015$ and 
  $\rm 12 + log(O/H) =  8.72$ (Allende Prieto, Lambert \& Asplund 2001). Fig 20 shows the O3N2 nebular metallicity averaged in radial annuli, using only pixels with $\rm F(H\alpha)>10^{-16}$ and $\rm signal/noise>3$ for both ratios (so there is a gap between the disk and the N hotspot). Pixels with $\rm N2>0.7$ or $\rm O3>0.7$ are also excluded on the basis that they are likely to be more dominated by shocks than by star-formation, but this has little effect on the mean O3N2 values (shocks might cause O/H to be underestimated by increasing the O3 ratio but this may largely be compensated by the increase in N2).  Fig 20 also shows the stellar lwZ and mwZ in the same annuli (using all pixels).

 The PP04 calibration gives a much higher metallicity than Marino et al. (2013), $12 + \rm log(O/H) \simeq 8.79/8.85$ integrated/central (1.17--$\rm 1.35 Z_{\odot}$), compared 
 to $\rm 8.57/8.62$ (0.71--$\rm 0.79Z_{\odot})$. The latter is much closer to the model-fit stellar mwZ. From $r=0.5$ to $r=11.5$ arcsec the Marino et al. O3N2 metallicity decreases by $0.080\pm 0.006$ dex, but beyond this remains flat (perhaps even increasing slightly in the outer hotspot).
  A linear fit at $r<12$ arcsec gives $8.6166-0.0069(\pm 0.0002)r$ (plotted on Fig 20), and for $r<8$ arcsec only, $8.6205-0.0083(\pm 0.0003)r$. These gradients are -0.028 and -0.033 dex/kpc, or with
  the disk exponential scale $h=6.5$ arcsec (Puerari et al 2005) corresponding to $r_{eff}=10.9$ arcsec (2.75 kpc), can also be expressed as  -0.076 and  -0.091 dex/$r_{eff}$.
   \begin{figure}
 \includegraphics[width=0.72\hsize,angle=-90]{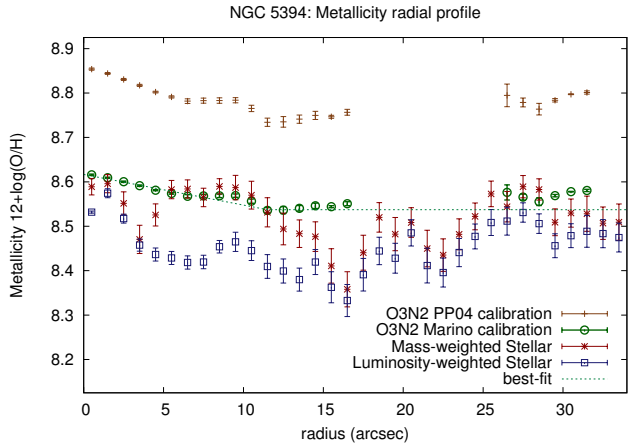}
\caption{Radial profiles of the nebular (O3N2) metallicities with PP04 and Marino et al. (2013) calibrations (with our fitted gradient), and of the luminosity-weighted and mass-weighted stellar metallicities from the {\sevensize PORTO3D} fits.} 
 \end{figure}
 
   S\'anchez et al. (2014) found CALIFA disk galaxies typically had a metallicity gradient of -0.1 dex/$r_{eff}$ at $0.3<r< 2~r_{eff}$, with flattening at even larger radii and evidence of a shallower mean gradient (-0.05) for interacting than for non-interacting (-0.11) subsamples (these estimates used the PP04 calibration and would be reduced by 1/3 for that of Marino et al.). Rich et al. (2012) found similar gradients, which flattened as mergers progressed. NGC 5394 appears consistent with their mean for early-stage mergers, but also with the non-interacting disks of the same mass in Ho et al. (2015). 
   Simulations predict that interactions reduce the central O/H  and flatten gradients  by the infall of  low-metallicity gas (Rupke, Kewley \& Barnes 2010), but there is also a competing effect of re-enrichment by star-formation  (Torrey et al. 2012), so the resulting evolution is more complex (in disk galaxies with high gas fractions the central O/H may even be increased).
   
     A new study by 
   Barrera-Ballesteros et al. (2015), comparing interacting and isolated disk galaxies in CALIFA,  suggests that the effect of interactions on O/H is more complex than a simple progressive flattening. They found interactions had little effect on the average central O/H, and might even steepen metallicity gradients at small radii ($r<0.5~r_{eff}$) where interacting galaxies show an enhancement in SFR. At larger radii ($r\sim 1$--$2~r_{eff}$)  the mean metallicity is reduced and it must be here the gradient is flattened. This was attributed to the inflow of lower metallicity gas. Also gradients in the outer galaxy could  be `stretched' out by the formation of tidal tails (Rich et al. 2012). These processes could explain NGC 5394 having a `normal' O/H gradient at $r<10$ arcsec which becomes flat only at  $r> r_{eff}$.

  The mwZ shows no significant gradient in the disk (from $r=0$ to $r=8.5$ arcsec changing by $-0.0018\pm 0.0331$), whereas the lwZ falls steeply ($-0.1135\pm0.0151$  between $r=0.5$ and $r=6.5$), and remains much lower than the mwZ in the outer disk, e.g. by $0.1359\pm 0.0222$ at $r=8.5$ (a $6\sigma$ difference). This is surprising in view of the expectation that metallicity would increase with time, which would give $\rm lwZ>mwZ$. Again a possible explanation is that the intermediate age and recent starbursts were fuelled by inflowing gas of lower metallicity compared to the underlying older disk stars, and, as Fig 8 of  Barrera-Ballesteros et al. (2015) suggests, the resulting O/H reduction is most apparent not at the centre but at $r\sim 
 5$--10 arcsec (0.5--1~$r_{eff}$).
 
    \begin{figure}
 \includegraphics[width=0.72\hsize,angle=-90]{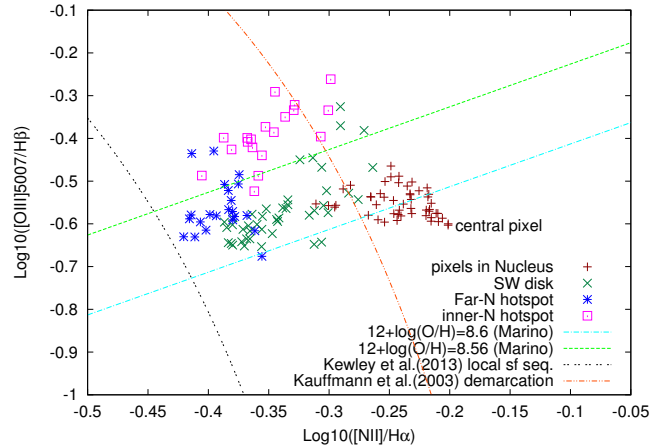}
\caption{`BPT' diagram plot of the log of O3 vs. N2, for the pixels in the regions of strong star-formation, showing the local star-forming sequence from Kewley et al. (2013), the HII/composite demarcation of Kauffmann et al. (2003) and two iso-metallicity lines based on the Marino et al. (2013) calibration.}
\end{figure}

  Returning to the nebular metallicity, Fig 21 shows the BPT (Baldwin, Phillips \& Terlevich 1981) diagram, i.e. log O3 vs. log N2, for the principal star-forming regions, together with two isometallicity lines (using the Marino et al. calibration), and the `local star-forming sequence' from Kewley et al. (2013), which is obtained from a photoionization model (with low electron density) and fits the locus of typical $z<0.1$ star-forming galaxies in the SDSS. The latter is essentially a sequence of increasing metallicity (right/downwards) and approximately orthogonal to the isometallicity lines [on which $\rm \Delta(log~N2)\simeq \Delta(log~O3)$].
  The distribution of CALIFA pixels shows the increase in metallicity from the outer star-forming regions to the central pixel, but combined with this, a large shift in the orthogonal direction, along the isometallicity lines and away from the local sequence, by as much as 0.2 dex in N2.  This latter cannot be due to metallicity but is as expected for a steep increase in $n_e$ towards the centre (Kewley et al. 2013) and we propose this is the cause.
   The entire nuclear region transgresses the Kauffmann et al. (2003) demarcation, which was intended to divide pure star-forming galaxies from HII/AGN composites, but is too conservative in not fully allowing for the potential [NII] enhancement of high $n_e$ and  hard radiation fields, which can move HII regions across the line (e.g. Fig 2 of  S\'anchez et al. 2015).

 \section{Kinematics}
 \subsection{Rotation of the Disk}
    \begin{figure}
 \includegraphics[width=0.82\hsize,angle=-90]{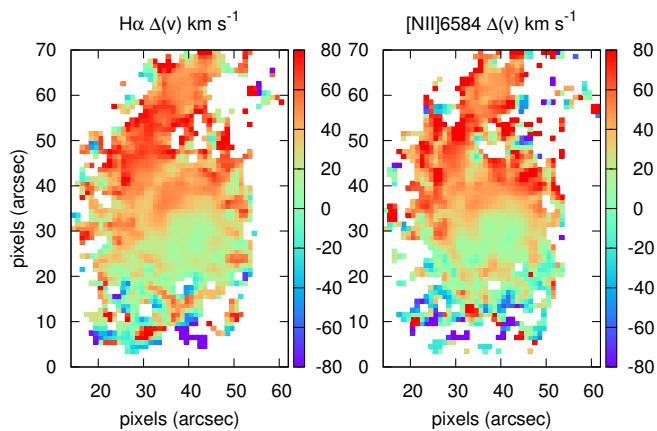}
\caption{Maps of the line-of-sight relative velocity $\Delta(v)$, from the red/blue-shifting of the emission lines $\rm H\alpha$ (left) and [NII]6584 (pixels shown white, which have little emission, do not have valid measurements). Note they are closely similar.} 
 \end{figure}
 
   \begin{figure}
 \includegraphics[width=0.73\hsize,angle=-90]{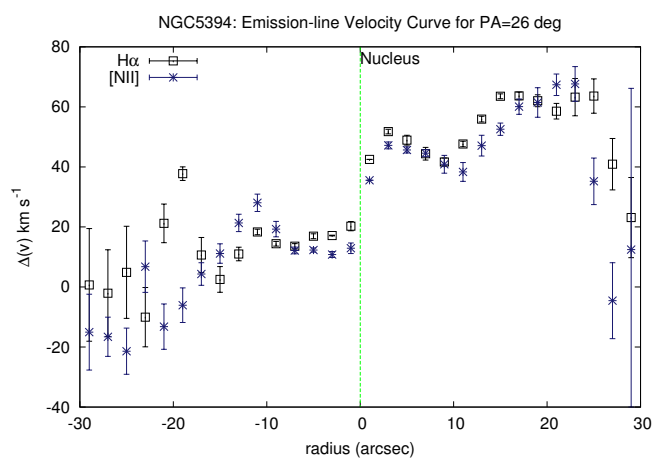}
\caption{Line-of-sight velocity curve from the $\rm H\alpha$ wavelength averaged in
$\pm 15^{\circ}$ sectors on the kinematically determined axis, position angle $\phi_k=26^{\circ}$ (East of North), with the NNE direction denoted positive.}
 \end{figure}
 
    \begin{figure}
 \includegraphics[width=0.82\hsize,angle=-90]{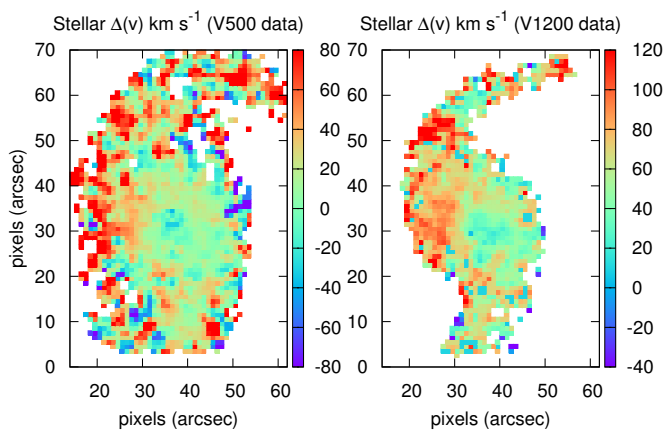}
\caption{Map of the stellar line-of-sight relative velocity $\Delta(v)$ measured by fitting to the spectra (i.e. from absorption lines) on the V500 and the V1200 data.} 
 \end{figure}
 
    \begin{figure}
 \includegraphics[width=0.73\hsize,angle=-90]{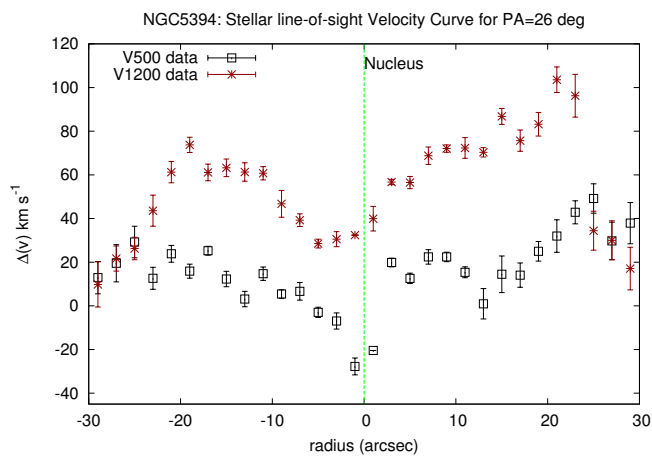}
\caption{Line-of-sight velocity curve from the stellar spectra, shown for both the V500 and V1200 data averaged in $\pm 15^{\circ}$ sectors on the position angle $\phi_k=26^{\circ}$.}
 \end{figure}

 Fig 22 shows a map of the line-of-sight velocity as measured from the $\rm H\alpha$ line, and also for [NII]6584. The galaxy is  inclined only $15^{\circ}$ from face-on, according to K02, so that the line-of-sight $\Delta(v)$ is lower than the true rotation of the disk   by a factor 3.9. Even so, the velocity gradient from rotation is clearly visible in the $\rm H\alpha$ velocity map. We obtain a velocity curve $v_{rot}(r)$ by averaging this map in two opposite sectors centred on the nucleus, in 2-arcsec bins of radius and within position angles   $\theta=\phi_k\pm 15^{\circ}$ and $ \theta=(\phi_k+\pi)\pm 15^{\circ}$ where $\phi_k$ is the kinematic position angle of the galaxy. The velocity curve is evaluated in this way for all PAs $\phi$ and  $\phi_k$ estimated as the angle which maximises the velocity difference between the opposite sides, i.e. $\Sigma_{r=0}^{r=22"} [v_{rot}(r)-v_{rot}(-r)]$;  this gave $\phi_k=26\pm13^{\circ}$, in  agreement with the kinematic  (receding side) PA of $25.8^{\circ}$ measured for this galaxy by 
  Garc\'ia-Lorenzo et al. (2015).

We take $\phi_k$ as $26^{\circ}$ and Fig 23 shows the $\rm H\alpha$ $v_{rot}(r)$ along this axis (NNE is positive, SSW is negative). The velocity curve shows the gradient produced by the galaxy's rotation, with an amplitude (taken as half the difference between the mean of the two largest velocity measurements on each side) 35 km $\rm s^{-1}$. It is relatively symmetric, but with minima on both sides at $r\simeq 10$ arcsec followed by  steeper gradients, a feature seen in some barred galaxies and which might correspond to the weak distortion K99 found in the $\rm H\alpha$  velocity map some 8 arcsec NE of the nucleus.  The [NII] velocity curve is very similar and gives an amplitude 
 43 km $\rm s^{-1}$. The disk rotation velocity must be at least $\sim 40\times 3.9=156$  km $\rm s^{-1}$,  which is typical for a spiral galaxy in this  stellar mass range ($\sim 2.4$--$4.6\times 10^{10}M_{\odot}$), e.g. Cortese et al. (2014).

 Shocks and ionized outflows are associated with mergers (Rich et al. 2011; Rich, Kewley \& Dopita 2014),  and might be revealed by kinematics, e.g. Yuan et al. (2012) attributed a high-N2 region to shocks/outflows as it was $\sim 100$ km $\rm s^{-1}$ blueshifted with respect to the galaxy. The $\Delta(v)$ map for [NII]6584 might be more  sensitive to shocks than $\rm H\alpha$. However,  the two velocity maps are very similar, and neither shows any region of blueshifted emission (ionized outflowing gas), either at the nucleus or in the outer (annular) high-N2 region.

 Fig 24 shows the stellar  $\Delta(v)$ maps, obtained by fitting the emission-subtracted spectra with {\sevensize STARLIGHT} stellar models, for both the V500 data and the higher-resolution V1200, and Fig 25 the stellar velocity curves measured in the same way on the PA $\phi_k=26^{\circ}$, centring on the nucleus in each case. The stellar $\Delta(v)$ maps, especially from V500, show  a localized blueshift ($\sim 20$ km $\rm s^{-1}$) near $r\simeq 0$ and the cause is discussed below.
The V500 stellar velocities show a weaker velocity gradient, with an amplitude (excluding the centre) only 21 km $\rm s^{-1}$, but those from the V1200 data have a stronger gradient and a positive offset with maximum amplitude 49 km $\rm s^{-1}$.  Cortese et al. (2014) found that stellar rotation velocities in SAMI (Sydney-Australian-Astronomical-Observatory Multi-object Integral-Field Spectrograph survey) disk galaxies averaged 0.14 ($\pm 0.11$) dex lower than $v_{rot}$ from nebular emission lines, but considering the large uncertainties and low inclination it is not clear if the two differ significantly in this galaxy. 

\subsection{Evidence of an Outflow}
 We did not see evidence of outflowing ionized gas in the $\rm H\alpha$ and [NII] velocity maps. Also both line profiles appear quite symmetric (although asymmetry was reported by K99), at least at the resolution of the V500 data, and are well-fitted by Gaussians. Any blueshifted line emission must be very faint compared with the star formation is the disk. However, the  cool gas phase of outflows may instead show itself in absorption lines, in particular MgII(2796,2803) and NaI (5890,5896). The nuclear spectrum plotted in Fig 4 has a very strong absorption line at $5890\rm \AA$ restframe, which can be identified as the NaI (5890,5896) doublet, unresolved by the V500 grating and slightly blueshifted. While this line appears in purely stellar spectra (especially from K type stars), a stronger and blueshifted  NaI absorption, seen in many LIRGs (e.g. Heckman et al. 2000), is a signature of starburst-driven outflows.    
 
 To investigate this we sum the stellar model fitted by {\sevensize PORTO3D} over the same 5 pixels of the nucleus and compare this with the observed spectrum (Fig 26). The NaI line observed here with $\lambda=5889.62\pm0.09\rm \AA$  and $\rm EW=7.03\pm0.06$  is $>3$ times stronger than, and blueshifted with respect to, the purely stellar model where the blended doublet is fitted by a Gaussian with $\lambda =5892.73\pm 0.24\rm \AA$ with $\rm EW=2.26\pm 0.07\AA$. The observed line blueshift is $3.11\pm 0.26\rm \AA$ or  $158\pm 13$ km $\rm s^{-1}$.  Taking the (observed-model) residual spectrum isolates the non-stellar, interstellar-gas component of the NaI line, which is here $\lambda=5888.45\pm 0.09\rm\AA$, and by comparison with the ({\sevensize STARLIGHT}-model) stellar NaI gives the blueshift of the non-stellar component as $4.38\pm 0.26\rm \AA$ or  $223\pm13$  km $\rm s^{-1}$. 
   
 This is indicative of a neutral gas outflow from the nucleus, directed perpendicular to the disk. In Fig 7, the spectrum of the nucleus region, integrated over 50 pixels ($r<4$ arcsec), shows enhanced NaI absorption  with $\rm EW=5.52\pm 0.02\AA$ at $\rm \lambda=5890.12\pm 0.02\AA$ (almost as strong as for the central 5 pixels only), but elsewhere the NaI line is consistent with the stellar models in EW (only 1--$2.5\rm \AA$) and $\lambda(\simeq 5892$--$3\rm \AA$). This confirms the strong blueshifted absorption originates in the galaxy centre and is not a spurious feature produced by sky lines or some instrument problem. 
   The outflow appears to have a similar, but weaker effect on the  CaII `K' absorption line at $3933\rm\AA$; in the central 5 pixels the observed and stellar model wavelengths are measured as $3931.48\rm\AA$ (EW $3.6\rm\AA$), $3933.04\rm\AA$ (EW $2.0\rm\AA$), and the residual (excess absorption) is at $3930.30\rm\AA$, which corresponds to a very similar outflow blueshift $\Delta(v)=-209$ km $\rm s^{-1}$.   This outflow might also explain the stellar velocity maps appearing slightly blueshifted at the centre. These $\Delta(v)$  could have been locally biased (up to $\sim 20$ km $\rm s^{-1}$ blueward) by the effect of the outflow on some absorption lines, especially in the V500 spectra which contain the NaI doublet.

 We calculate (integrating over wavelength) a map of the NaI EW (Fig 27); the regions of excess absorption approximately fills the central $r<4$ arcsec ($r<1$ kpc) and, measuring with {\sevensize IRAF} ellipse, has a FWHM  6.4--7.0 arcsec (much larger than the instrumental FWHM of 3.5 arcsec) and in the outer isophotes is slightly extended to the SE (at the $\rm EW=3.5\AA$ isophote, $\rm PA=-36\pm 3$ deg and ellipticity$=0.24\pm0.02$; using {\sevensize IRAF} ellipse). Alongside we show a $\Delta(v)$  map obtained by Gaussian-fitting the NaI absorption in each pixel (with {\sevensize IRAF} fitprof, over 5879 to $5920\rm\AA$, avoiding the weak HeI emission line at $5876\rm\AA$).  The blueshifted and high-EW regions correspond spatially, being extended by $> 1.5$ kpc on a  PA close to the inner-disk minor axis, which seems consistent with an outflow perpendicular to the disk.  Over this central region the total flux absorbed in the residual (non-stellar i.e. outflow) line is $1.44\times 10^{-14}$ erg $\rm cm^{-2}s^{-1}$.

          \begin{figure}
 \includegraphics[width=0.73\hsize,angle=-90]{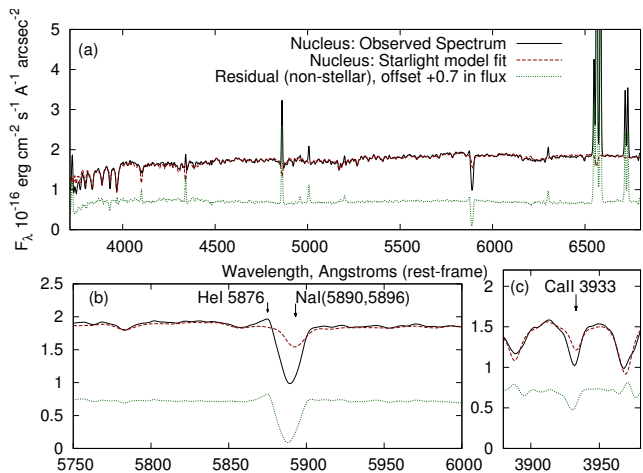}
\caption{Observed spectrum of the central 5 pixels compared with the {\sevensize STARLIGHT} stellar model fit, for (a) the whole spectrum and (b) for the region of the NaI(5890,5896) absorption doublet, together with the
 observed - model residual, showing that there is substantial blueshifted, non-stellar NaI absorption at the central region of NGC 5394. (c) This blueshifted absorption may also be visible, less strongly, in the CaII (K) line at $3933\rm\AA$.} 
 \end{figure}

      \begin{figure}
 \includegraphics[width=0.88\hsize,angle=-90]{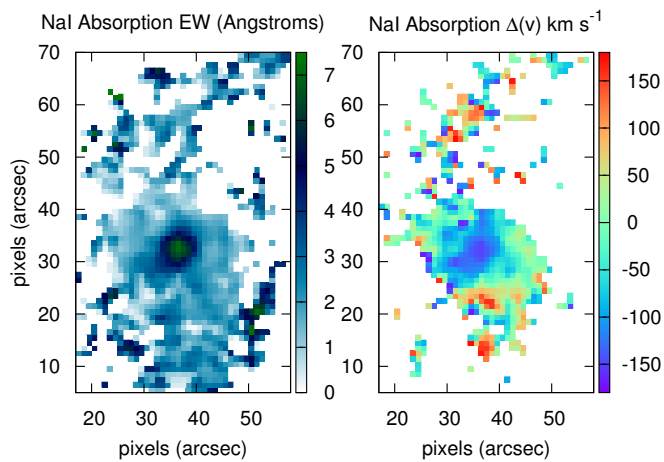}
\caption{(a) Map of the equivalent width of the NaI(5890,5896) absorption doublet, obtained by integrating over the line (5879--$5904\AA$), clearly showing the central region of strong absorption (up to $\rm EW=7\AA$), in excess of that expected for stars alone (about $2\rm\AA$); (b) map of the velocity shift $\Delta(v)$ of the NaI absorption line relative to the stellar-fit $5892.73\rm\AA$,  as obtained by a Gaussian fit to each pixel over 5879--$\rm 5920\AA$.} 
\end{figure}

 The EW and $\Delta(v)$  of the NaI absorption are comparable to the NaI lines in many of the starbursting, $\rm L_{IR}=10^{10-12}L_{\odot}$ galaxies in Heckman et al. (2000), which have dust-rich super-winds. Blueshifted NaI absorption is observed to a lesser degree in more typical disk galaxies (Chen et al. 2010); the detection probability increases with central SFR, face-on orientation and high dust reddening, and the correlation of $\Delta(v)$  with disk orientation implies the  outflows are perpendicular and entrained within bicones of opening angles $\leq 60$ degrees (Wild et al. 2014 measured this angle directly for  NGC 4676A as $\Omega_w/4\pi=0.37$, corresponding to $51^{\circ}$). The NaI blueshift/outflow velocity is also correlated with SFR, with $\sim 200$ km $\rm s^{-1}$ for $\rm 10~M_{\odot}yr^{-1}$, and Bordoloi et al. (2014) found the same correlations for $z\sim1$ galaxy  outflows observed in MgII 2796,2803.  Our $\Delta(v)$ is slightly above average for the SFR but within the typical range.
  
 Heckman et al. (2000) give a formula (eq 5) for estimating outflow rate as
 $\rm{dM/dt}\sim 60(r_*/kpc)(N_H/3\times 10^{21} cm^{-2})\times(\Delta {\it v}/200~km s^{-1})\times
 (\Omega_{\it w}/4\pi) M_{\odot}yr^{-1}.$
  Putting in starburst radius $r_{*}=0.75$ kpc, $N_H=6.1\times 10^{20}$ at the centre (from Fig 5 of K02), $\Delta v=223$ km $\rm s^{-1}$ and $\Omega_w/4\pi\simeq 0.37$ , gives  $\rm {dM/dt}\simeq 3.77~M_{\odot}yr^{-1}$, similar to our estimate for the SFR. Heckman et al. (2000, and similarly Bordoloi et al. 2014) concluded that typically a `superwind is expelling matter at a rate comparable to the star formation rate'. The kinetic power would be $\rm {1\over2}{dM\over dt}v^2= 5.9\times 10^{40}$ erg $\rm s^{-1}$; 1/13 the galaxy's dust-corrected $\rm H\alpha$ luminosity  (Table 1).

\subsection{Evidence of Inflow?}
The outer high-excitation region (at $9<r<16$ arcsec) does not show the same blueshifted $\Delta(v)$ as the outflow from the nucleus. Although very apparent in line ratio maps it seems to leave no imprint on the velocity maps in either $\rm H\alpha$ or [NII], suggesting little motion perpendicular to the disk of  the underlying galaxy. Nor is it associated with a significant broadening of the [NII] line (at least at this resolution), which remains at $\rm FWHM\simeq 6\AA$ across the galaxy. This may be  similar to the late-stage merger  NGC 3256, in which the outer regions with high, shock-like line ratios are blueshifted only `a few 10 km $\rm s^{-1}$'  (Rich et al. 2011), whereas the central strong NaI absorption (caused by outflow) is blueshifted by $\Delta(v)=-309$ km $\rm s^{-1}$ (Heckman et al. 2000).

The  non-visibility of the outer high-excitation region in $\Delta(v)$ maps would argue that it is not outflowing gas driven by the nuclear starburst. We suggest  that the starburst-driven outflow remains collimated in front of the central  disk, as on the NaI map (Fig 27), and the outer high excitation region is powered  instead by the kinetic energy of gas spiralling inwards within the disk plane, where it would have little $\Delta(v)$ in the line-of-sight direction, as a result of tidal forces from the interaction. The gas would lose angular momentum in the shocks and eventually form stars.  

This mechanism was argued for the outer regions of high excitation/line ratios found in a number of interacting or merging LIRGs (Monreal-Ibero et al. 2010, Rich et al. 2011).  One argument was that the occurrence of outer high-excitation regions was more correlated with (late) interaction stages than with high SFR.
 In the Torrey et al. (2012) merger simulations for a similar mass galaxy, the nuclear gas inflow rate is generally $\sim 0.5$--$\rm 1.0\times SFR$ and 1--5 $\rm M_{\odot} yr^{-1}$. At the velocities of disk rotation $\sim 160$--170 km $\rm s^{-1}$ or as estimated for the shocks, $\sim 250 $ km $\rm s^{-1}$, and an inflow rate of at a few $\rm M_{\odot}yr^{-1}$,  the kinetic energy input would be sufficient to power the shock component of line emission ($\sim 10^{40}$ erg $\rm s^{-1}$).

 \section{History of the Interaction}
 NGC 5394 is now undergoing a nuclear starburst (generating an outflow), has other localized knots of star-formation, and shows evidence of a more extensive (over most of the galaxy) episode of star-formation $10^8$ to $10^9$ years ago. Fig 28 shows the 
 the {\sevensize PORTO3D} model-fits mass distributions of young ($<100$ Myr) and intermediate age (0.1-1.0 Gyr) stars.  The young population has a total mass 206 million $\rm M_{\odot}$ and is very centrally concentrated with 132 million $\rm M_{\odot}$ in the central $<4$ arcsec and only 1.7 million in the far-N hotspot. The dust-corrected $\rm H\alpha$ SFR in Table 1, especially if further corrected to the Salpeter IMF of these masses (giving 6 $\rm M_{\odot}yr^{-1}$) implies the current SFR is significantly higher than the mean ($\rm 2.06~M_{\odot}yr^{-1}$) for the past $10^8$ yr. Similarly our SFR of $\rm 3.39~M_{\odot}$ is more than twice the $1.54~\rm M_{\odot} yr^{-1}$ estimate of Lanz et al. (2013) based on SED-fitting with a model of the same IMF and 100 Myr of constant star-formation.
$\rm H\alpha$ traces only the very recent SFR ($\leq 10$ Myr), compared to longer timescales for the infra-red emission, so the implication is again the SFR has increased strongly over the past $10^8$ yr.
   \begin{figure}
 \includegraphics[width=0.86\hsize,angle=-90]{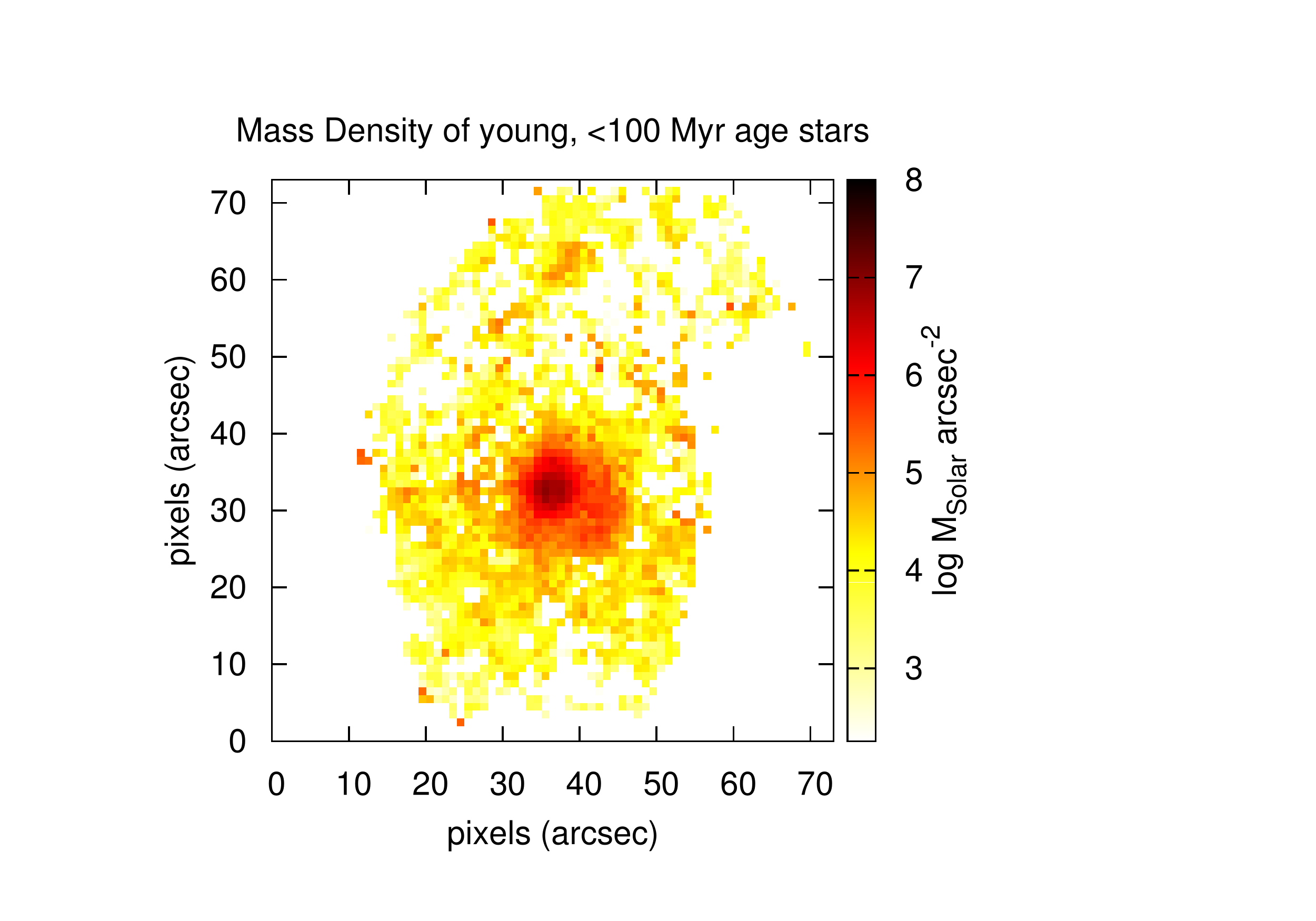}
  \includegraphics[width=0.86\hsize,angle=-90]{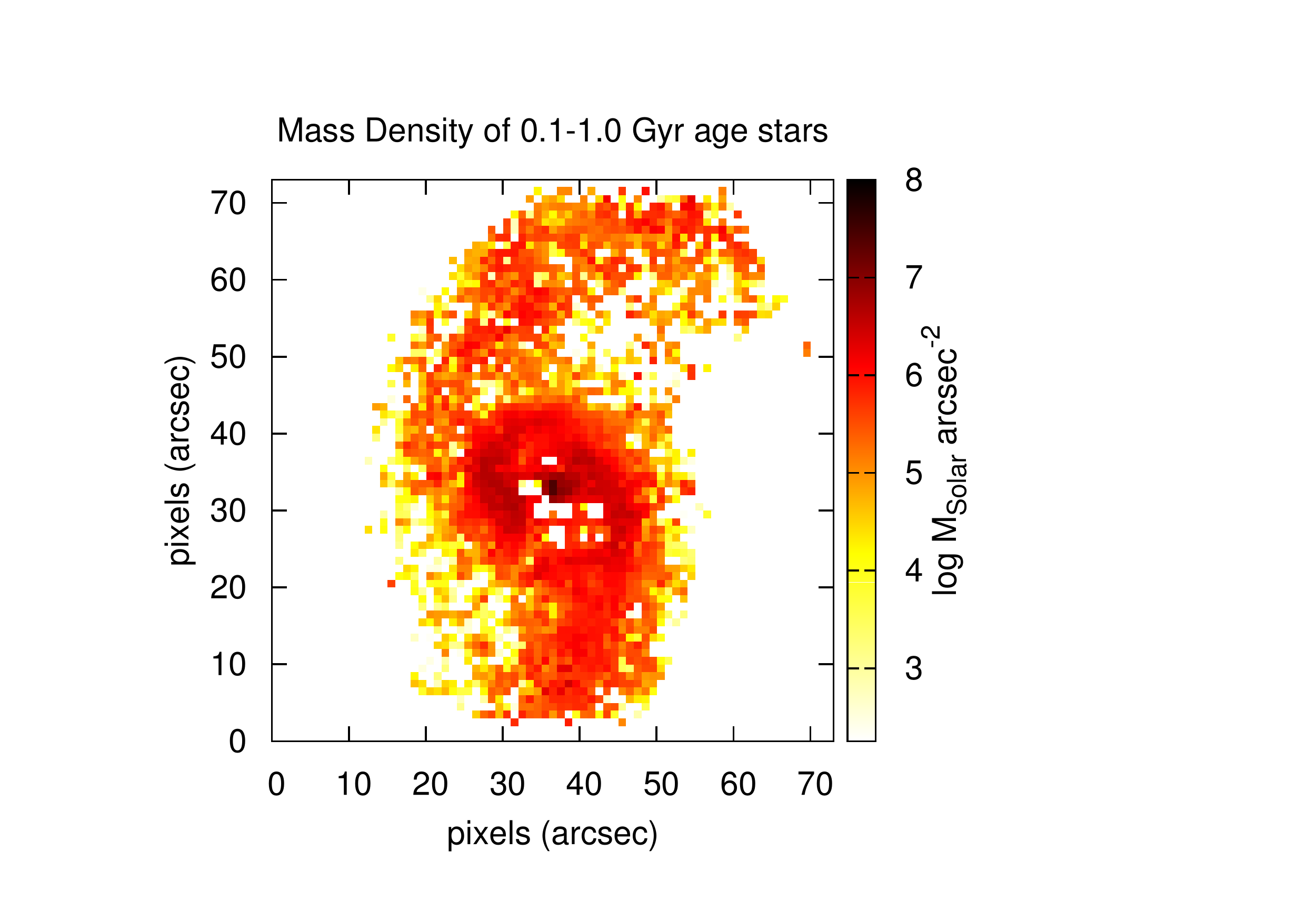}
\caption{Distribution of the young (age zero to 100 Myr) and  intermediate age stellar population (age 100 Myr to 1 Gyr), as estimated from the {\sevensize STARLIGHT} model fits, shown on a log scale of stellar mass per pixel.} 
 \end{figure}

 The intermediate-age population has a very different distribution, in the outer disk as a broad ring and, especially,  throughout the outer tidal arms. The earlier starburst must have been spread over the whole galaxy, whatever its shape at that time.   The total mass of intermediate age stars is estimated as  $\rm 1.200\times 10^9M_{\odot}$ with only 170 million $\rm M_{\odot}$ at $r<4$ arcsec. These patterns recall the merger simulation of Renaud, Bournaud \& Duc (2015), in which early star-formation is extensive but the final burst is `almost exclusively central, due to gas inflows'.
    The concentration of star-formation in the SW disk (the inner western arm, see K02), reflects the lopsided distribution of $\rm H_2$ which would be a short-term effect of the interaction, so is seen in the young population map but not for the intermediate-age stars, which instead are concentrated in the other two disk arms (eastern and outer-western). This agrees with to the `west-vs.-east time lag' in star- and arm-formation (K99), with only the inner-western `third arm' that is delayed i.e. recent.  Similarly the two  hotspots are not seen in the intermediate-age map, as they are younger features.

  K99/K02 describe   NGC 5394 as `post-ocular' and similar to the interacting galaxy IC 2163 but at a later stage of evolution, with star-formation more concentrated in the nucleus.
 IC 2163 is interacting with the larger spiral NGC 2207 and indeed appears eye-shaped with most of its star-formation in numerous hotspots forming two arcs around the rim of the `eye' (Kaufman et al. 2012). In the  Struck et al. (2005) simulation of this pair there was a major starburst at the first close passage, some 280 Myr earlier, and there is predicted to be a second burst at least 100 Myr in the future as the galaxies finally merge. NGC 5394 could be 50--100 Myr ahead of IC 2163 and entering this stage, with the SFR increasing again as the galaxy approaches merger.  
In this simulation the  long tidal arms are drawn out of the initially round galaxy very soon after the first close encounter and  starburst, and this might account for the tidal arms in NGC 5394 having the highest content of intermediate-age stars. 
 
 The Arp 84 system is certainly at a later evolutionary/merger stage than the Mice, for which  the simulations of Wild et al. (2014) put the first perigalacticon only 170 Myr ago, and observationally the oldest star clusters in both galaxies date from exactly that time (Chien et al. 2007). The Mice have less $\rm H\delta$ absorption and much less of an intermediate age population than NGC 5394, and in the simulations of Wild et al. (2014) and Barnes (2004) their biggest starburst lies 0.4--0.8 Gyr in the future when the galaxies will merge and the SFR peak at 40--$\rm 50 M_{\odot} yr^{-1}$.

 The hotspots in the northern arm of NGC 5394 are examples of the `hinge clumps' seen in many interacting galaxies (Smith et al. 2014) --  bright knots of active star-formation, typically found where tidal arms/tails join to the disk, with $\rm EW(H\alpha)\sim 100\AA$ and $L(\rm H\alpha)\sim 10^{40}$ erg $\rm s^{-1}$. Analytical and numerical models of tidal disturbance (Struck and Smith 2012, SS12) can explain many of the galaxy's features including the formation of long, curved and paired tidal arms (which can extend to $\sim 3$ times the radius of the pre-interaction galaxy) and an inner disk with a structure described as ocular with internal arms (e.g. Figs 7 and 10 of SS12); furthermore the model predicts `swallowtail caustics'  at the base of the tidal tails or arms where intersecting streams of gas would pile up, giving
  rise to localised star-formation which could be quite sustained and  produce  `hinge clumps'  such as the inner-N hotspot.  A refinement of the SS12 model, where a prograde encounter is represented by successive impulses, can produce additional swallowtail caustics  far out from the disk on the spiral arms (Fig 12 of SS12) and in this way account for the formation and position of the far-North hotspot (and the similar feature in Arp 82). Introducing a little asymmetry further improves the resemblance to real Arp (1966) galaxies.
   The high-resolution simulation of Renaud et al. (2015) also predicts the formation of a `young massive cluster' far out on the northern tidal arm. Probably these hotspots will become globular clusters.

 In a further attempt to reconstruct the star-formation history we run {\sevensize STARLIGHT} model fits on the 7 region spectra (Fig 7), with a set of Chabrier-IMF stellar  templates representing 23 ages from 1 Myr to 13 Gyr and four metallicities (Z=0.004, 0.008, 0.02, 0.05). For each $\rm f_{\lambda}(age,Z)$ template spectrum the program fits the mass of stars present now and at formation. As the results are rather noisy, mean SFRs are estimated for wide  $\rm \Delta(log[age])=0.3$ bins, by summing in each the (at-formation) masses of stars formed, and dividing by the bin width in years. In this model the current stellar mass is only $\rm 2.44\times 10^{10}M_{\odot}$, reflecting the `lightweight' IMF.
The model also fits dust extinction to the stellar continuum, $A_v= 1.06$ mag for the nuclear region (notably less than the emission-line $A_{\rm H\alpha}=2.32$ mag from the Balmer decrement), and  $A_v=0.60$ and 0.31 for the SW and outer disk.

Fig 29 shows the resulting SFR histories. The nucleus has a large old ($\sim 10$ Gyr) stellar population, plus resurgent star-formation (note increasing SFR) over the past 100 Myr. The outer disk and tidal arms are dominated by intermediate age ($10^9$--$10^8$ yr old) stars, the $\rm H\alpha$-bright SW disk region has additional star-formation at age $<100$ Myr, the star-formation in the far-N hotspot also goes back some 100 Myr, while the inner-N hotspot, with stronger Balmer absorption, is mostly  intermediate-age. The model fitted a large component of $\sim 0.9$  Gyr age stars for the outer disk and tidal arms, suggesting the interaction with NGC 5395 and enhanced star-formation have been going on for a long time, perhaps even with two previous close passages.
  \begin{figure}
 \includegraphics[width=0.7\hsize,angle=-90]{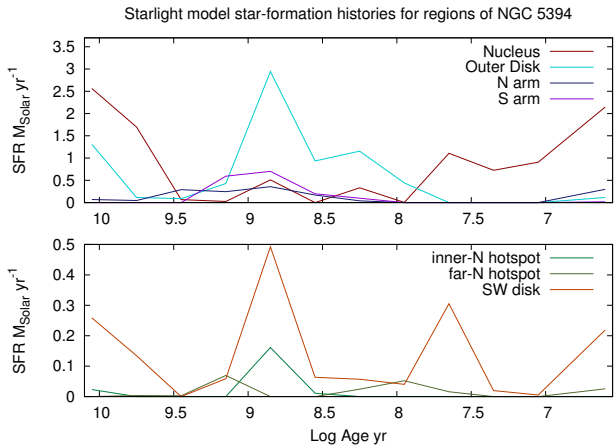}
\caption{Reconstructed SFR history, estimated from the Starburst model fits, of large (above) and and small (below) regions of NGC 5394 as a function of look-back time, plotted in $\rm \Delta(log[age])=0.3$ intervals.} 
 \end{figure}

 The first perigalacticon could have triggered a prolonged influx of gas towards the nucleus, fuelling the subsequent star-formation. If this gas was relatively low metallicity, the stellar metallicity would then decrease during the interaction,  which (we argued in Section 5) could account for the (model-fit) 
 luminosity-weighted stellar metallicity being lower than the mass-weighted (Fig 20).  Yet this influx did not flatten the O3N2-metallicity gradient at $r\leq 10$ arcsec, perhaps because of competing re-enrichment by star-formation (Torrey et al. 2012). Also O3N2 could be less sensitive to mixing of low-Z gas if prolonged star-formation and reprocessing produced a high central N/O ratio (e.g. Belfiore et al. 2015). 
 Tidally-induced gas infall may be be the cause and the energy source of the relatively faint (a few percent of the total line emission) but extensive ($\sim 8$ kpc diameter) outer ring of elevated N2, O3, $\rm[OI]/H\alpha$ and $\rm [SII]/H\alpha$ ratios, which are consistent with shocks (of velocity $\sim 250$ km $\rm s^{-1}$). Extensive shock  emission features  similar to this this have been found in other late-stage or strongly interacting, gas-rich galaxies  (Monreal-Ibero et al. 2010, Rich et al. 2011, 2014; Rodr\'iguez-Baras et al. 2014).

 We also find evidence for a nuclear-starburst-driven  outflow of gas with (blueshift) velocity  $223\pm 13$ km $\rm s^{-1}$, and estimated the rate as $\sim  3.77~ \rm M_{\odot}yr^{-1}$, which from Heckman et al. (2000) might not be exceptional if the SFR is near our estimate of 3.39 $\rm M_{\odot} yr^{-1}$.
 Wild et al. (2014) estimated  8--$\rm 20~M_{\odot}yr^{-1}$ at   350 km $\rm s^{-1}$ for the ionized outflow from NGC 4676A. That the outflow in NGC 5394 is seen as absorption in the neutral line Na I, rather than in emission, might be a feature of the later evolutionary stage, with the neutral outflow persisting longer, as suggested by the detection of  neutral outflows in some dusty, red post-starburst galaxies (e.g. Sato et al. 2009), and in recent gas-rich mergers (e.g. NGC 34 in  Schweizer \& Seitzer 2007).
  Much higher spectral resolutions ($R\geq 5000$)  may be required  to  measure velocity dispersions and disentangle the different inflow, outflow or rotating components.

 Another feature of NGC 5394 is the high electron density of $n_e\sim 750$ $\rm cm^{-3}$ in the central kpc,  compared to the central $n_e\sim 200$--$\rm 300~cm^{-3}$ more typical of star-forming galaxies. This is not unique, e.g. Krabbe et al. (2014), in a Gemini Multi-Object Spectrograph study of 12 galaxies in 7 interacting pairs, found even lower [SII] 6717/6731 ratios and higher $n_e$ ($\sim 1000$ $\rm cm^{-3}$) in some examples,
 (such as AM 1054-325B and AN 1219-430), and 
  concluded that there was evidence interacting galaxies on average had higher $n_e$ but, on the basis of moderate $\rm [OI]6300/H\alpha$ and $\rm [OIII]5007/H\beta$ ratios, this was not generally caused by shock excitation. Otherwise no cause was given except to attribute a central increase in $n_e$ to gas inflow.
  The high central $n_e$ of NGC 5394 could be a consequence of sustained gas infall over hundreds of Myr. Perhaps the $n_e$ enhancement is maximised for  this particular interaction geometry (coplanar, prograde, with larger companion) and stage. Large integral field surveys which can measure line ratios diagnostic of $n_e$ will be able to quantify how it correlates with local and galaxy-scale properties.

 \section{Conclusions}
 
(i) We examine CALIFA data for the spiral galaxy NGC 5394, which is strongly interacting   with the $4\times$ more massive spiral NGC 5395.
Most of the $\rm H\alpha$ emission from NGC 5394, and therefore the star-forming activity,  is concentrated in the nucleus ($75\%$ in  the central $r\leq 1$ kpc)  with additional star-forming regions in the SW of the disk and in two hotspots on the northern tidal arm. The galaxy is dusty, especially in the central $r\leq1$ kpc where the $\rm H\alpha/H\beta$ ratio corresponds to a $\rm H\alpha$ extinction of 2.32 magnitudes. For the whole galaxy we estimate a dust-corrected, $\rm H\alpha$ based star-formation rate of 3.39 $\rm M_{\odot}yr^{-1}$ (Chabrier IMF). The outer northern-arm hotspot has only $0.9\%$ of the total SFR but the highest $\rm H\alpha$ equivalent width anywhere in the galaxy, over $100\rm \AA$. Isolated, extranuclear concentrations of star-formation, like this hotspot, are predicted to form in merger simulations.
 
(ii) Much of the galaxy outside the inner disk has a post-starburst spectrum with strong Balmer absorption ($\rm  EW_{abs}( H\delta)\geq 6 \AA$).  The $\rm H\delta$ map and spectral model-fitting show the galaxy underwent an extensive episode of rapid star-formation 0.2--1.0 Gyr ago. Comparison with simulations and other mergers suggest this was triggered by the earlier close passage of the two galaxies.

(iii) NGC 5394 has a very high (for a purely star-forming galaxy) $\rm [NII]/H\alpha$ ratio of 0.63 at its centre. The $\rm [OIII]/H\beta$, $\rm [OI]6300/H\alpha$ and $\rm [SII]/H\alpha$ ratios remain low here, with no central peak, and thus disfavour AGN and shocks as explanations. Instead the enhanced central N2 ratio is probably due to increased collisional excitation associated with the very high central electron density, $n_e\simeq 750~\rm cm^{-3}$ on the basis of the [SII]6717/6731 ratio. 

(iv) There is an outer region, forming an irregular ring of pixels around the nucleus at $9<r<16$ arcsec ($2.25<r<4$ kpc), with even higher $\rm [NII]/H\alpha$  ratios of 0.65--0.85, on average 0.74. Here the  $\rm [OIII]/H\beta$, $\rm [OI]6300/H\alpha$ and $\rm [SII]/H\alpha$ ratios are also high, suggesting a substantial ($\simeq 50\%$ of $\rm H\alpha$) contribution from shocks (we estimate, with $v\sim 250$ km$\rm s^{-1}$ and no pre-ionization).

(v) The O3N2 index of nebular metallicity with the lower, Marino et al. (2013) calibration and the mass-weighted model-fit stellar metallicity are similar at $\rm 12 + log(O/H)\simeq 8.6$ (0.76 solar). The metallicity from the O3N2 index shows a negative radial gradient of about -0.03 dex/kpc (typical of disk galaxies) at $r<10$ arcsec (or one $r_{hl}$), but this flattens at larger radii. Also the luminosity-weighted stellar metallicity is lower than the mass-weighted, especially in the outer disk,  which might be the result of earlier inflow of lower metallicity gas.

(vi) The NaI(5890,5896) absorption in the spectrum of the nucleus has an equivalent width 6 to $7\rm \AA$, at least 3 times stronger than expected from stellar absorption alone, and is blueshifted by $\rm 3\AA$ over a $>1.5$ kpc region. The residuals between the observed line and a stellar model indicate the non-stellar absorption component is  blueshifted by 223 km $\rm s^{-1}$. This is evidence for a neutral outflow of gas, at this velocity and perpendicular to the disk plane, driven by the central starburst.

(vii) Although the galaxy is viewed only $\sim 15^{\circ}$ deg from face-on, a velocity gradient is visible in the $\rm H\alpha$ and [NII] emission lines, and in the stellar spectra (at least for the V1200 data), consistent with a disk rotation velocity $\sim 156$--180 km $\rm s^{-1}$ (as expected from the stellar mass). The $\rm H\alpha$ and [NII] velocity maps are very similar and show no signature of an ionized outflow (such as a blueshifted region) either at the nucleus or the outer ring of high N2 ratios. This suggests that shocks in the outer high-excitation region are not produced by the central outflow but more likely by interaction-triggered gas inflow within the disk plane.

(viii) The strong neutral gas outflow, high electron density and concentrated star formation ( $>1$ $\rm M_{\odot}yr^{-1}kpc^{-2}$) at the nucleus, and the strong Balmer absorption and hotspots elsewhere, are all a consequence of the prolonged interaction with NGC 5395. Over a timescale as long as  $\sim 1$ Gyr and one or more perigalactica, this has reshaped the galaxy, formed tidal arms, triggered starbursts and gas inflow, and now the SFR may be increasing again as the two galaxies approach merger. 

\section*{Acknowledgments}
This study uses data provided by the Calar Alto Legacy Integral Field Area (CALIFA) survey (http://califa.caha.es/), funded by the Spanish Ministery of Science under grant ICTS-2009-10, and the Centro Astron\'omico Hispano-Alem\'an. Based on observations collected at the Centro Astron\'omico Hispano Alem\'an (CAHA) at Calar Alto, operated jointly by the Max-Planck-Institut f\"ur Astronomie and the Instituto de Astrof\'isica de Andaluc\'ia (CSIC). 

NR, AH, JMG, PP and PL acknowledge Funda\c{c}\~{a}o para a Ci\^{e}ncia e a
Tecnolog\'ia (FCT) support through UID/FIS/04434/2013, and through project
FCOMP-01-0124-FEDER-029170 (Reference FCT PTDC/FIS-AST/3214/2012) funded by
FCT-MEC (PIDDAC) and FEDER (COMPETE), in addition to FP7 project
PIRSES-GA-2013-612701.
NR acknowledges the support of  FCT postdoctoral grant SFRH/BI/52155/2013 followed by CAUP2014-04UnI-BPD (University of Porto). 
 AH also acknowledges a Marie Curie Fellowship
co-funded by the FP7 and the FCT (DFRH/WIIA/57/2011) and FP7 / FCT
Complementary Support grant SFRH/BI/52155/2013.
 PL is supported by a Postdoctoral grant SFRH/BPD/72308/2010, funded by the FCT. JMG acknowledges support by the FCT through the Fellowship SFRH/BPD/66958/2009
and POPH/FSE (EC) by FEDER funding through the program Programa Operacional
de Factores de Competitividade (COMPETE).
PP is supported by FCT through the Investigador FCT Contract No. IF/01220/2013 and 
POPH/FSE (EC) by FEDER
funding through the program COMPETE.

\section*{References} 

\vskip0.15cm \noindent Allen M.G., Groves B., Dopita M., Sutherland R., Kewley L.J., 2008, ApJS 178, 20.

\vskip0.15cm \noindent Allende Prieto C., Lambert D.L., Asplund M., 2001, ApJ 556, L63.
	
\vskip0.15cm \noindent Arp H., 1966, ApJS 14, 1..

\vskip0.15cm \noindent Arp H., 1969, A\&A 3, 418.

\vskip0.15cm \noindent Baldwin J.A., Phillips M.M., Terlevich R., 1981, PASP 93, 5.

\vskip0.15cm \noindent Barnes J.E., 2004, MNRAS, 350, 798.

\vskip0.15cm \noindent Barrera-Ballesteros J.K., et al., 2015, A\&A 579, 45.

\vskip0.15cm \noindent Belfiore F., et al. 2015, MNRAS 449, 867.

\vskip0.15cm \noindent Bordoloi R., et al., 2014, ApJ 794, 130.

\vskip0.15cm \noindent Brinchmann J., Pettini M., Charlot, S., 2008,  MNRAS 385, 769. 

\vskip0.15cm \noindent Calzetti D., Armus L., Bohlin R.C., Kinney A.L., Koornneef J., Storchi-Bergmann T., 2000, ApJ 533, 682.

\vskip0.15cm \noindent Cant\'o J., Elliott K.H., Meaburn J., Theokas A.C., 1980, MNRAS 193, 911. 

\vskip0.15cm \noindent Casoli F., et al., 1998, A\&A, 331, 451. 

\vskip0.15cm \noindent Chen Yan-Mei, Tremonti C.A., Heckman T.M., Kauffmann G., Weiner B.J., Brinchmann J., Wang J., 2010, AJ 140, 445.

\vskip0.15cm \noindent Chien Li-Hsin, Barnes J.E., Kewley L.J., Chambers K.C., 2007, ApJ 660, 105.

\vskip0.15cm \noindent Cid Fernandes R., et al.  2013, A\&A 557, 86.

\vskip0.15cm \noindent Cortese L., et al., 2014, ApJ 795, 37.

\vskip0.15cm \noindent Farage C.L., McGregor P.J., Dopita M.A., Bicknell G.V., 2010, ApJ 724, 267.

\vskip0.15cm \noindent Garc\'ia-Benito R., et al., 2015, A\&A, 576, 135.

\vskip0.15cm \noindent Garc\'ia-Lorenzo B., et al. 2015, A\&A, 573, 59.

\vskip0.15cm \noindent Gomes J.M., et al., 2015, A\&A, submitted.

\vskip0.15cm \noindent Goto T., 2008, MNRAS 381, 187.

\vskip0.15cm \noindent Heckman T.M., Armus L., Weaver K.A., Wang J., 1999, ApJ 517, 130.

\vskip0.15cm \noindent Heckman T.M., Lehnert M.D., Strickland D.K., Armus L.,  2000, ApJS, 129, 493.
 
\vskip0.15cm \noindent  Higdon S.J.U, Higdon J. L.,Smith B. J.,Hancock M.,  2014, ApJ 787, 103.

\vskip0.15cm \noindent  Ho I-Ting, Kudritzki R-P, Kewley L.J., Jabran Zahid H., Dopita M.A.,  Bresolin F., Rupke. D.S.N., 2015, MNRAS 448, 2030.

\vskip0.15cm \noindent  Howell J.H., et al., 2010, ApJ 715, 572.

\vskip0.15cm \noindent Husemann B., et al., 2013, A\&A 549, 87

\vskip0.15cm \noindent Kaneko H., Kuno N., Iono D., Tamura Y., Tosaki T.; Nakanishi K., Sawada T.,  2013, PASJ 65, 20.

\vskip0.15cm \noindent Kauffmann, G. et al. 2003, MNRAS, 346, 1055

\vskip0.15cm \noindent Kaufman M., Brinks E., Elmegreen B.G., Elmegreen D.M.,; Klari\'c  M., Struck C., Thomasson M., Vogel S., 1999, AJ 118, 1577.

\vskip0.15cm \noindent  Kaufman M., Sheth K., Struck C.; Elmegreen B.G., Thomasson M., Elmegreen D. M., Brinks E., 2002, AJ 123, 702.

\vskip0.15cm \noindent Kaufman M., Grupe D., Elmegreen B.G. Elmegreen D.M., Struck C., Brinks E., 2012, AJ 144, 156.

\vskip0.15cm \noindent Keel W.C., Kennicutt Jr. R.C., Hummel E., van der Hulst J.M., 1985, AJ 90, 108.

\vskip0.15cm \noindent Kennicutt Jr. R. C., 1998, ARA\&A 36, 189.

\vskip0.15cm \noindent Kewley L.J., Dopita M.A., Leitherer C., Dav\'e R., Yuan T., Allen M., Groves B., Sutherland R., 2013, ApJ, 774, 100.

\vskip0.15cm \noindent Krabbe A.C., Rosa D.A., Dors O.L., Pastoriza M.G., Winge C., H\"agele G.F., Cardaci M.V., Rodrigues I., 2014, MNRAS 437, 1155.

\vskip0.15cm \noindent Lanz L., et al., 2013, ApJ 768, 90.

\vskip0.15cm \noindent  Le Tiran L., Lehnert M.D., van Driel W., Nesvadba N.P.H., Di Matteo P.,
 2011, A\&A 534, 4.
 
 \vskip0.15cm \noindent  Lehmer B.D., Alexander D.M., Bauer F.E., Brandt W.N., Goulding A.D.,  Jenkins L.P., 
 Ptak A., Roberts T.P., 2010, ApJ, 724, 559.

\vskip0.15cm \noindent L\'opez-Hern\'andez J., Terlevich E., Terlevich R., Rosa-Gonz\'alez D., D'az ç.; Garc\'ia-Benito R.,;V\'ilchez J., H\"agele G., 2013, MNRAS 430, 472.
 
\vskip0.15cm \noindent Marino R.A., et al.,  2013, A\&A 559, 114.

\vskip0.15cm \noindent Monreal-Ibero A., Arribas S., Colina L., Rodr'guez-Zaur\'in J., Alonso-Herrero A., Garc\'ia-Mar\'in M., 2010, A\&A 517, 28.

 \vskip0.15cm \noindent Papaderos P., et al., 2013 A\&A 555, 1.

\vskip0.15cm \noindent Pettini M., Pagel B.E.J. 2004, MNRAS, 348, L59.

\vskip0.15cm \noindent Puerari I., Valdez-Guti\'errez M., Hern\'andez-L\'opez I., 2005, AJ 130, 1524.

\vskip0.15cm \noindent Renaud F., Bournaud F., Duc P-A., 2015, MNRAS, 466, 2038.

\vskip0.15cm \noindent Rich J.A., Kewley L.J., Dopita M.A., 2011, ApJ, 734, 87.

\vskip0.15cm \noindent  Rich J.A., Kewley L.J., Dopita M.A., 2014, ApJ 781, L12.

\vskip0.15cm \noindent Rich, J.A., Torrey P., Kewley L.J.,  Dopita M.A., Rupke D.S.N., 2012, ApJ, 753, 5.
	
\vskip0.15cm \noindent Rodr\'iguez-Baras M., Rosales-Ortega F.F., D\'iaz A.I., S\'anchez S.F., Pasquali A., 2014, MNRAS 442, 495.

\vskip0.15cm \noindent Rupke D.S.N., Kewley L.J., Barnes J.E., 2010, ApJ 710, 156.

\vskip 0.15cm  \noindent  S\'anchez S.F., et al. 2012, A\&A 538, 8.

\vskip 0.15cm  \noindent  S\'anchez S.F., et al. 2014, A\&A 563, 49.

\vskip 0.15cm  \noindent  S\'anchez S.F., et al. 2015, A\&A 574, 47.

\vskip0.15cm \noindent S\'anchez-Blazquez P., et al. 2014, A\&A 570, 6.
 
 \vskip0.15cm \noindent Sato T., Martin C.L., Noeske K.G., Koo D.C., Lotz J.M., 2009, ApJ 696, 214.

 \vskip0.15cm \noindent Schweizer F., Seitzer P., 2007, AJ 133, 2132.
 
\vskip0.15cm \noindent Sobral D., Best P.N., Smail I., Mobasher B., Stott J., Nisbet D., 2014, 
MNRAS 437, 3516.
 
\vskip0.15cm \noindent  Struck C., Kaufman M., Brinks E., Thomasson M.,  Elmegreen B.G., Elmegreen D.M., 2005, MNRAS 364, 69.

\vskip0.15cm \noindent Struck C., Smith, B.J., 2012,  MNRAS, 422, 2444.

\vskip 0.15cm  \noindent  Smith, B.J., Soria R., Struck C., Giroux M.L., Swartz D. A.; Yukita M., 2014, AJ 147, 60.

\vskip 0.15cm  \noindent Smith B.J., Struck C., Hancock M., Appleton P.N., Charmandaris V., Reach W.T.,  2007, AJ 133, 791.

\vskip0.15cm \noindent Torrey P., Cox T.J., Kewley L., Hernquist L.,	 2012, ApJ 746, 108.

\vskip0.15cm \noindent Walcher C J., et al. 2014, A\&A 569, 1.

\vskip0.15cm \noindent  Wild V.; Groves B., Heckman T., Sonnentrucker P., Armus L., Schiminovich D., Johnson B., Martins L., Lamassa S., 2011, MNRAS 410, 1593.

 \vskip 0.15cm  \noindent Wild V., Rosales-Ortega F., Falcon-Barroso J, et al., 2014, A\&A 567, 143.
 
\vskip0.15cm \noindent Yuan T.-T., Kewley L.J., Swinbank A.M., Richard J., 2012, ApJ 759, 66.
 
 \end{document}